\documentclass[12pt,preprint]{aastex}
\usepackage{graphicx,natbib,amsmath}
\numberwithin{equation}{section}

\usepackage{lscape}
\newcommand{\lapprox} {\, \lower3pt\hbox{$\sim$}\llap{\raise2pt\hbox{$<$}}\,}
\newcommand{\gapprox} {\, \lower3pt\hbox{$\sim$}\llap{\raise2pt\hbox{$>$}}\,}

\begin{document}

\title{A CLASSIFICATION SCHEME FOR TURBULENT ACCELERATION PROCESSES IN SOLAR FLARES}

\author{Nicolas Bian\altaffilmark{1},
        A. Gordon Emslie\altaffilmark{2}, and
        Eduard P. Kontar\altaffilmark{1}}

\altaffiltext{1}{School of Physics \& Astronomy, The University, Glasgow G12 8QQ, Scotland, UK, n.bian@physics.gla.ac.uk, eduard@astro.gla.ac.uk}

\altaffiltext{2}{Department of Physics \& Astronomy, Western Kentucky University, Bowling Green, KY 42101; emslieg@wku.edu}

\begin{abstract}
We establish a classification scheme for stochastic acceleration models involving low-frequency plasma turbulence in a strongly magnetized plasma. This classification takes into account both the properties of the accelerating electromagnetic field, and the nature of the transport of charged particles in the acceleration region. We group the acceleration processes as either resonant, non-resonant or resonant-broadened, depending on whether the particle motion is free-streaming along the magnetic field, diffusive or a combination of the two. Stochastic acceleration by moving magnetic mirrors and adiabatic compressions are addressed as illustrative examples. We obtain expressions for the momentum-dependent diffusion coefficient $D(p)$, both for general forms of the accelerating force and for the situation when the electromagnetic force is wave-like, with a specified dispersion relation $\omega=\omega(k)$. Finally, for models considered, we calculate the energy-dependent acceleration time, a quantity that can be directly compared with observations of the time profile of the radiation field produced by the accelerated particles, such as during solar flares.

\end{abstract}

\keywords{Sun: activity -- Sun: flares -- Sun: particle emission -- Sun: X-rays, gamma rays}

\section{Introduction}

The acceleration of particles to non-thermal energies is a fundamental phenomenon throughout the Universe. In particular, the impulsive phase of a solar flare is characterized by the acceleration of charged particles, including electrons, protons and heavier ion species \citep[see, e.g.,][and references therein]{2011SSRv..159..107H,2011SSRv..159..301K,2011SSRv..159..357Z}. At the fundamental level, of course, charged particles are accelerated by an electric field; it is the degree of spatio-temporal coherence of this electric field that distinguishes acceleration mechanisms.

The action of an accelerating field can be divided into two main categories, depending on whether it relies on {\it systematic} or {\it stochastic} change in the energy of the particles. Systematic change of energy is involved in a a broad range of particle acceleration mechanisms developed over the years. In the solar flare context, these include the action of electric fields, either large-scale sub-Dreicer fields  \citep[e.g.,][]{1959PhRv..115..238D,1966AZh....43..340S,1969ARA&A...7..149S,1971ApJ...165..655F,1973SoPh...30..459A,1975AZh....52..316P,1977A&A....55...23H,1981A&A...103..331K,ben,2003PhPl...10.2732K,2005MNRAS.356.1107Z,2007A&A...472..623P}
or super-Dreicer fields in compact current sheets \citep[e.g.,][]{1969ARA&A...7..149S,1986A&A...163..210S,1988SoPh..115..149T,li1,li2,2010A&A...520A.105B},
and acceleration in collapsing magnetic traps \citep{1990JGR....9514925G,som1,som2,karl,2009A&A...508.1461G}.

On the other hand, when the energy change has a fluctuating character, the mechanism of particle acceleration is said to be {\it stochastic}. The classic example is second-order Fermi acceleration \citep{1949PhRv...75.1169F}, which is based on a more random motion of the scattering centers and/or of the particles. The mechanism which creates the corresponding distribution of accelerating impulses may involve both plasma waves \citep[e.g.,][]{mel,as,1994ApJ...434..747P,miller,1998A&A...331.1066R,2005PhPl...12j2110C,2006SSRv..124..277G,2010A&A...519A.114B,2012A&A...539A..43K,2012ApJ...749...58M}, magnetohydrodynamic turbulence \citep[e.g.,][]{1993PhyU...36.1020B,2008SSRv..134..207P,2009ApJ...692L..45B,2012ApJ...748...33C},
or turbulent electric fields in fragmented current sheets \citep[e.g.,][]{an,arz,car,turkmanietal05,2009A&A...508..993B,2011ApJ...729..101G}.

Extended work on particle acceleration in solar flares has shown that stochastic acceleration models \citep[e.g.,][]{1966PhFl....9.2377K,1966SvPhU...9..370T,1967PhFl...10.2620H,1969Ap&SS...4..143M,1975MNRAS.172..557S,1974ARA&A..12...71W,1979ApJ...233..383B,1992JPhG...18.1089J,1993JPlPh..49...63S}
not only avoid some of the fundamental large-scale electrodynamic issues characteristic of systemic acceleration models \citep{1995ApJ...446..371E, miller}, but also can produce an acceleration efficiency \citep{2008AIPC.1039....3E} that is broadly consistent with that deduced from hard X-ray observations. Given, then, the likelihood that stochastic acceleration plays an important role in the acceleration of non-thermal particles in solar flares, we seek in this work to synthesize various stochastic acceleration models in a unified manner.  In this way, we highlight their similarities and differences, and we show how their essential parameters could be constrained through observation of, e.g., turbulent and directed plasma motions \citep{1982SoPh...78..107A,1989ApJ...344..991F} or of magnetic fluctuations \citep{2011ApJ...730L..22K, 2011A&A...535A..18B} in solar flares.

We classify stochastic acceleration models by the nature of the spatial transport of particles in the acceleration region. We do this through the \citet{cor} approximation, a procedure for relating Eulerian and Lagrangian correlation functions which enjoys widespread use in the theory of turbulent transport \citep[see, e.g.,][]{1995PhRvE..51.4844W,2004PPCF...46.1051V,2006PhRvE..73b6404N,tau}. This approach allows us to divide stochastic acceleration models into three primary classes: resonant, resonant-broadening and non-resonant, depending on the nature of the transport process affecting the accelerated particles. The necessary analysis is presented both through an ordinary space representation and in the Fourier domain, a representation particularly appropriate when the acceleration mechanism involves waves. For each case we derive the form of the diffusion coefficient $D(p)$ (when it exists) as a function of the particle momentum $p$. We then use this to obtain the characteristic acceleration time as a function of particle energy $E$ in the non-relativistic regime.

We also present examples of scenarios that depend on the specific nature of the force accelerating the particles, with the turbulent fluctuations being associated either with weak wave turbulence or strong turbulence. It is emphasized that stochastic acceleration can also occur when the accelerating field is coherent, provided the particle transport in this field is itself stochastic, for instance diffusive.

In Section~\ref{stoc_general}, we review the fundamental concepts of stochastic acceleration mechanisms through a discussion of the \citet{ob} model, one of the first mathematical formulations of turbulent acceleration. In Section~\ref{acc_scenarios}, we derive the essential features of resonant, non-resonant, and resonant-broadened models, respectively. Examples of different forms of the accelerating force are considered in Section~\ref{forms}. A brief discussion of the energy spectra that result is given in Section 5 and an overall summary is presented in Section~\ref{summary_sec}.

\section{Stochastic Acceleration}\label{stoc_general}

\subsection{Fundamental Concepts}\label{concepts}

\subsubsection{Velocity and Spatial Diffusion}\label{diffusion_eqs}

Consider a particle of momentum $p$ subjected to a rapidly fluctuating force $F(t)$, so that the (one-dimensional) motion of the particle is described by a set of Langevin equations

\begin{equation}\label{obukhov}
\dot{p}=F(t) \,\, ; \qquad m\dot{x} = p \equiv mv \,\,\, .
\end{equation}
The random force $F(t)$ is taken to be a Gaussian process with zero average $<F(t)> \, =0$ and non-zero variance $<F^2>$.  The force $F(t)$ is delta-correlated in time:

\begin{equation}
C(t,t')\equiv \, <F(t) \, F(t')> \, =2D \, \delta(t-t') \,\,\, ,
\end{equation}
where $D$ is a diffusion coefficient and the averaging is made over ensemble realizations. As a consequence of stationarity, the (auto-) correlation function has the property that $C(t,t')=C(t-t')$.

The \citet{ob} model~(\ref{obukhov}) is the undamped version of the Ornstein-Uhlenbeck process, described by

\begin{equation}\label{ou}
\dot{p}=F(t)- \frac{p}{\tau} \,\, \, ; \qquad m\dot{x}=p \,\,\, ,
\end{equation}
where $\tau $ is a damping time. From the corresponding Fokker-Planck equation for the distribution $P(p,t)$ of momentum at time $t$, one finds that the stationary distribution has a Gaussian form

\begin{equation}\label{maxwell}
P(p)= \frac{1}{\sqrt{2\pi <p^2>}}\exp\left(-\frac{p^2}{2<p^2>}\right) \,\,\, ,
\end{equation}
where $<p^2>=\tau D$ is the variance characterizing the momentum distribution. The mean square displacement grows linearly with time, $<x^{2}>=2D_{x}t$, where the spatial diffusion coefficient $D_x$ is related to the momentum damping rate by $D_{x}=\tau(<p^{2}>/m^{2})$. Spatial diffusion occurs only for thermalized particles when the statistics of the process have become stationary, i.e., when $<p(t) \, p(t')>$ depends only on $(t-t')$ according to $<p(t) \, p(t')> \, = \, <p^2> \exp(-\mid t-t'\mid/\tau)$.

On the other hand, the undamped ($\tau \rightarrow \infty $) Obukhov model results instead in {\it momentum} diffusion: the mean energy of the accelerated particles, $<p^2>$, grows linearly with time. The equation describing the evolution of the probability density $P(x,p,t) \, \equiv  \, <\delta [x-x(t)] \, \delta[p-p(t)]>$ is the Fokker-Planck equation

\begin{equation}\label{fpp}
\frac{\partial P(x,p,t)}{\partial t}+\frac{p}{m} \, \frac{\partial P(x,p,t)}{\partial x}= D \, \frac{\partial^2 P(x,p,t)}{\partial p^2} \,\,\, .
\end{equation}
With the initial condition $P(x,p,t=0)=\delta(x) \, \delta(p)$, the solution of this equation is a Gaussian form exhibiting simultaneous unbounded spreading with respect to both momentum and position and with the following scaling behaviors \citep{benf}:

\begin{equation}
<p^{2}>=2Dt \,\,\, ; \qquad <x^{2}>={\frac{2Dt^{3}}{ 3 m^2}} \,\,\, .
\end{equation}
Averaged over position $x$, the probability density $P(p,t)$ is a Gaussian with a variance that grows linearly with time. Similarly, averaged over momentum $p$, the probability density $P(x,t)$ is a Gaussian with a variance that has a cubic dependence on time\footnote{The dispersions in velocity and in space are related by $<p^{2}> \, \propto \, <x^{2}>^{1/3}$, a result that can be used to deduce the Kolmorogov law $E(k)\propto k^{-5/3}$ of hydrodynamic turbulence. This was the original motivation for the \citet{ob} model. Later, it was used by \citet{dup} in his foundation of resonance broadening theory applied to wave-particle interactions in plasmas.}.

It should be emphasized that the standard diffusion law $<p^{2}> \, \propto t$ produced by this model is a particular case. In general, stochastic acceleration processes result in a momentum diffusion characterized by $<p^{2}> \, \propto t^{\alpha}$, with $\alpha\neq 1$. Cases with $0<\alpha<1$ are termed {\it sub-diffusive} and cases with $\alpha>1$ are termed {\it super-diffusive} \citep{2000EJPh...21..279B}. {\it Anomalous} diffusion occurs when the momentum diffusion coefficient is zero, infinite, or if it depends on momentum. In general, the momentum diffusion coefficient is momentum-dependent when (a) the force itself depends on momentum, or (b) the force depends on position as well as on time.

\subsubsection{Spatially-Varying Force Fields and Correlation Times}\label{spat_var}

We now consider a non-trivial extension of the Obukhov model which incorporates an explicit spatial dependence of the force $F(x,t)$:

\begin{equation}\label{s1}
\dot{p}=F(x,t) \,\,\, , \qquad m\dot{x}=p \,\,\, .
\end{equation}
The force is still assumed to be homogeneous and stationary, with the statistical property $<F(x,t)> \, =0$.  Its {\it Eulerian correlation function} is

\begin{equation}
C(x,t) \, \equiv \, <F(0,0) \, F(x,t)>.
\end{equation}
We next define the Eulerian spatial and temporal scales by

\begin{equation}\label{taueuler}
\tau=\frac{1}{C(0,0)}\int_{0}^{\infty} dt \, C(0,t) \,\,\, ; \quad \lambda=\frac{1}{C(0,0)}\int_{-\infty}^{\infty} dx \, C(x,0) \,\,\, ,
\end{equation}
respectively, where

\begin{equation}\label{c00}
C(0,0) = \, <F^{2}> \,\,\, .
\end{equation}
A general expression for the Eulerian correlation function is therefore

\begin{equation}\label{cxtexp}
C(x,t) = \, <F^{2}> \, {\hat c} \left (\frac{x}{\lambda},\frac{t}{\tau} \right) \,\,\, ,
\end{equation}
where the non-dimensional function ${\hat c}$ satisfies the following conditions:

\begin{equation}\label{conditions}
{\hat c}(0,0)=1 \, ; \quad \int_{-\infty}^\infty {\hat c} \left ({\frac{x} { \lambda}}, 0 \right ) d \left ({\frac{x}{\lambda}} \right ) = 1 \, ; \quad \int_0^\infty {\hat c} \left (0, {\frac{t}{\tau}} \right ) d \left ({\frac{t}{\tau}} \right ) = 1 \,\,\, .
\end{equation}
A suitable function\footnote{Note the appearance of a factor of four in the temporal Gaussian exponent; this is a direct consequence of the half-space limits $[0,\infty)$ of the temporal integral (versus the full-space limits $(-\infty,\infty)$ in the spatial integral) in Equations~(\ref{taueuler}).} satisfying these properties is the Gaussian

\begin{equation}\label{ref_gaussian}
{\hat c} \left (\frac{x}{\lambda},\frac{t}{\tau} \right) = e^{-\pi x^2/\lambda^2} \, e^{-\pi t^2/4 \tau^2} \,\,\, .
\end{equation}

With the spatio-temporal statistics of the force now known, the problem is to calculate the momentum diffusion coefficient (when the latter exists) and to determine the form of the distribution function. \citet{taylor} established that the momentum diffusion coefficient $D(p)$ is related to the time integral of the {\it Lagrangian correlation function}, viz.

\begin{equation}\label{taylor_result}
D(p)=\int _{0}^{\infty} dt \, C_{L}(p \, ; t) \,\,\, ,
\end{equation}
where the Lagrangian correlation function of the force is defined by

\begin{equation}\label{lag_corr}
C_{L}(p \, ;t)=<F(0,0) \, F(x(t),t)> \,\,\, .
\end{equation}
Taylor's result (\ref{taylor_result}) follows from $<p^{2}> \, = \int _{0}^{t}dt'\int _{0}^{t} dt''<F(t')F(t'')> \, = \, 2\int _{0}^{t}dt' \, C_{L}(t') \, (t-t')$ and letting $t\rightarrow \infty$ (when the integral converges). The {\it Lagrangian correlation time} $\tau_{L}$, the correlation time associated with the force felt by a particle along its trajectory, is defined as

\begin{equation}\label{tauldef}
\tau_{L}=\frac{1}{C_{L}(0)}\int_{0}^{\infty} dt \, C_{L}(p;t) \,\,\, .
\end{equation}
Since $<F^{2}> \, = C(0,0) = C_{L}(t=0)$, Equations~(\ref{c00}), (\ref{taylor_result}) and~(\ref{tauldef}) show that the diffusion coefficient, when it exists, is given by

\begin{equation}\label{lagrangian_result}
D(p)=\tau_{L} \, <F^2> \,\,\, .
\end{equation}
The special cases $D=0$ (sub-diffusion) and $D \rightarrow \infty$ (super-diffusion), corresponding to $\tau_{L}$ being zero or infinity, respectively, have been discussed by \citet{bi} on the basis of a generalized Obukhov model.

In general, both $\tau_L$ and $<F^2>$ may depend on the particle momentum $p$, so that $D=D(p)$. We shall find that a typical dependence $D(p)$ is a power law

\begin{equation}\label{dprelation}
D(p)=D_{0} p^{\alpha} \,\,\, .
\end{equation}
Then, since $d<p^{2}>/dt=2D(p) = 2D_{0}p^{\alpha}$, it follows that

\begin{equation}\label{p_growth}
<p^{2}> \, \propto t^{2/(2-\alpha)} \,\,\, ,
\end{equation}
for $\alpha<2$.  For $\alpha > 2$, $<p^2>$ can attain infinite values within a finite period of time as may be seen by solving the equation $dp^2/dt = 2 D_0 p^\alpha$ with $\alpha > 2$.

It is clear from Equation~(\ref{lagrangian_result}) that determination of the diffusion coefficient essentially involves determination of the Lagrangian correlation time $\tau_{L}$. When the force is independent of position (as in the Obukhov model), determination of $\tau_{L}$ is a trivial task, because in this simple case $\tau_{L}=\tau$, the Eulerian correlation time. A much more significant challenge is to obtain the Lagrangian correlation scales (as a function of the Eulerian ones) when the force depends on position as well as time; indeed, it is this very spatial dependence of the force that makes the stochastic acceleration problem inherently non-linear. A general strategy can be summarized as follows: we seek the functional dependence of the Lagrangian correlation time $\tau_{L}$ (and hence $D$) on the Eulerian quantities $\tau$, $\lambda$ and possibly\footnote{\citet{dup} was the first to propose a scheme, based on the Obukhov model, to include the functional dependence of $\tau_{L}$ on $<F^{2}>$.} $<F^{2}>$. From dimensional arguments, if the time scale $\tau_{L}$ depends on the length scale $\lambda$ this dependence is expected to also involve quantities related to the spatial transport of particles, such as the particle momentum $p$ (if the particles are free-streaming) or the particle diffusivity coefficient $\kappa$ (if the transport of particles involve some form of spatial diffusion). The pertinent transport time scales are $m\lambda/p$ and $\lambda^{2}/\kappa$, respectively.

\subsubsection{Remarks}

It must be noted that the above analysis is one-dimensional.  Nevertheless, it is still a valid description of the behavior of a distribution of charged particles moving under the influence of random electric fields in a volume permeated by a uniform magnetic field $\mathbf{B}_{0}$, if the $x$-coordinate is taken to represent the coordinate parallel to the magnetic field and the accelerating force\footnote{Other possible types of force fields are discussed in Section 4.} is expressed as a parallel electric field: $F_x(t)=q E(x,t)$.  It is well known that the presence of a strong background magnetic field permits the decomposition of the motion of the charged particles into a guiding-center motion parallel to the $x$-axis and a much faster perpendicular gyration around the guiding magnetic field line. At this stage, we identify the position $x$ of the particle with that of the gyrocenter; this approximation will be relaxed later (Section~\ref{flr}).  The basic mechanisms for perpendicular acceleration of particles by transverse electric fields can also be outlined within the context of a one-dimensional model \citep{stu}.

In this work we restrict ourselves to the consideration of low-frequency turbulence and so do not consider the effect of gyroresonances.  Further, we employ a test-particle approach, so that effects of the collective behavior of the particles are neglected. The dissipation of electromagnetic energy resulting from the stochastic acceleration of particles will be discussed in a separate publication.

\subsection{The Corrsin approximation}\label{corrsin}

Let us write the Lagrangian correlation function~(\ref{lag_corr}) in the equivalent form

\begin{equation}\label{cldef}
C_{L}(p \, ;t)=\int dx<F(0,0) \, F(x,t) \, \delta [x-x(t)]> \,\,\, ,
\end{equation}
where $x(t)$ is a solution of the equations of motion (\ref{s1}). One way of obtaining a relation between the Lagrangian correlation $C_{L}(p;t)$ and the Eulerian correlation $C(x,t)$ is to invoke a procedure due to \citet{cor}, in which $x(t)$ is replaced by its statistical average, so that we may replace $\delta [x-x(t)]$ in Equation~(\ref{cldef}) by $<\delta [x-x(t)]>$.  This leads to the factorization

\begin{equation}
C_{L}(p \, ;t)=\int dx<F(0,0)F(x,t)> \, <\delta [x-x(t)]>
\end{equation}
and immediately allows us to write

\begin{equation}\label{cors}
C_{L}(p \, ;t)=\int dx \, C(x,t) \, P(x,t) \,\,\, ,
\end{equation}
where $P(x,t)\equiv <\delta [x-x(t)]>$ is the probability density for a particle to move from its starting point at the origin to position $x$ over time~$t$. Equation (\ref{cors}) shows that the Lagrangian correlation function is the spatial integral of the product of two quantities:

\begin{enumerate}

\item $C(x,t)$, the Eulerian correlation function, which depends on the properties of the force field $F(x,t)$,
and

\item the probability function $P(x,t)$, describing the spatial transport of particles in the acceleration region.

\end{enumerate}
This separation into properties of the accelerating force and properties of the particle trajectories forms the basis for our classification of acceleration models. Specifically, we shall categorize stochastic acceleration models primarily through the nature of the quantity $P(x,t)$, and then sub-categorize them according to the nature of the force $F(x,t)$ [and hence the Eulerian correlation $C(x,t)$] acting on the particles.

With the Lagrangian correlation function $C_L(p \, ;t)$ defined as in Equation~(\ref{cors}), the momentum diffusion coefficient $D(p)$ may be written

\begin{equation}\label{aa}
D(p)=\int_{0}^{\infty} dt \int_{-\infty}^\infty dx \, C(x,t) \, P(x,t) \,\,\, .
\end{equation}
At this stage, a preliminary comment can be made. First, let us use Equation~(\ref{cxtexp}) to write (\ref{aa}) in the equivalent form

\begin{equation}\label{aa1}
D(p) = \, <F^{2}>\int_{0}^{\infty} dt \int_{-\infty}^\infty dx \, {\hat c} \left ( \frac{x}{\lambda},\frac{t}{\tau} \right ) \, P(x,t) \,\,\, .
\end{equation}
We notice that when

\begin{equation}
P(x,t)\sim \delta(x) \,\,\, ,
\end{equation}
then

\begin{equation}
D(p) = \, <F^{2}> \, \int_{0}^{\infty} dt \, {\hat c} \left (0,\frac{t}{\tau} \right )=\tau <F^{2}> \, .
\end{equation}
This is essentially the Obukhov result $\tau_{L}=\tau$; the spatial dependence of the force, and hence the spatial transport of the particles, is irrelevant in evaluating the acceleration efficiency because the temporal variation of the force is taken to be infinitely fast, i.e., $\tau\rightarrow 0$. For such a situation, all stochastic acceleration models reduce to the common limit corresponding to the Obukhov model.

However, there are certainly situations where it is unreasonable to assume that the Eulerian correlation time $\tau \rightarrow 0$. An obvious example is when the force field acting on the particles is produced by wave motions, for which case the Eulerian correlation time $\tau$ is of the order of the wave period $T$. This in turn means that the Lagrangian correlation time $\tau_{L}$ now has a dependence on the Eulerian spatial scale $\lambda$, because the particles of necessity also feel the spatial variation of the force.

When acceleration results from the action of waves with a well-defined dispersion relation $\omega = \omega(k)$, it can be more convenient to express the momentum diffusion coefficient in terms of Fourier components of the Eulerian correlation function of the force field, i.e., in terms of the {\it spectrum} of the force field. The Fourier components ${\hat F}_{k,\omega}$ of $F(x,t)$ are defined through

\begin{equation}
F(x,t) = \sum_{k}\sum_{\omega} \, {\hat F}_{k,\omega} \,
e^{i(kx-\omega t)} \,\,\, ,
\end{equation}
with $k=n\delta k$, $\omega=m\delta \omega$, $\delta k=2\pi/L$, and $\delta \omega=2\pi/T$.  Here $L$ and $T$ are the spatial and temporal periods of the force field. Passing to the continuous case, the Fourier components ${\hat F}(k,\omega)$ of $F(x,t)$ are defined through

\begin{equation}
F(x,t) = \int \int dk \, d\omega \, {\hat F}(k,\omega) \, e^{i(kx-\omega t)} \,\,\, .
\end{equation}
Thus the Eulerian correlation function may be written in the form

$$C(x,x',t,t') \equiv \, <F(x,t)F(x',t')> \, =$$

\begin{equation}\label{def}
= \int \int \int \int dk \, dk'\, d\omega \, d\omega'\, <{\hat F}(k,\omega) \, {\hat F}(k',\omega')> \, e^{i(kx+k'x'-\omega t-\omega't')} \,\,\, .
\end{equation}
For the homogeneous and stationary case considered herein, the right-hand side depends only on the differences $x-x'$ and $t-t'$; this is possible only if $k'=-k$ and $\omega'=-\omega$.  This in turn means that the average $<{\hat F}(k,\omega) \, {\hat F}(k',\omega')>$ should be proportional to  $\delta(k+k') \, \delta(\omega+\omega')$:

\begin{equation}\label{sdefinition}
<{\hat F}(k,\omega) \, {\hat F}(k',\omega')> \, = S(k,\omega) \, \delta(k+k') \, \delta(\omega+\omega') \,\,\, ,
\end{equation}
which may be taken as a definition of the {\it force spectrum} $S(k,\omega)$.

The Eulerian correlation function $C(x,t)$ is thus the Fourier transform of the force spectrum:

\begin{equation}
C(x,t)=\int \int dk \, d\omega \, S(k,\omega) \, e^{i(kx-\omega t)} \,\,\, .
\end{equation}
Taking $x=0$ and $t=0$, we obtain the normalization

\begin{equation}\label{normf2}
C(0,0) = \, <F^{2}> \, = \int \int dk \, d\omega \, S(k,\omega) \,\,\, .
\end{equation}
As an example, for the Gaussian correlation function~(\ref{ref_gaussian}), the corresponding form of $S(k,\omega)$ is

\begin{equation}\label{skwgaussian}
S(k,\omega) = \left ( {\frac{\lambda \tau}{4 \pi}} \right ) \, <F^2> \, e^{-\lambda^2
k^2/4\pi} \, e^{- \omega^2 \tau^2/ \pi} \,\,\, .
\end{equation}
Under the Corrsin approximation, the momentum diffusion coefficient $D(p)$ (Equation~[\ref{aa}]) can also be expressed in terms of the force spectrum as

\begin{equation}\label{bb}
D(p) =  \int \int dk \, d\omega  \, S(k,\omega) \, G(k,\omega) \,\,\, ,
\end{equation}
or, in the discrete case, as

\begin{equation}\label{bbd}
D(p) =  \sum_{k}\sum_{\omega} \, |{\hat F}_{k,\omega}|^{2} \, G(k,\omega) \,\,\, .
\end{equation}
Here the function (which has the dimension of time)

\begin{equation}\label{prop_fourier}
G(k,\omega) = \int_0^\infty dt \, \int dx \, P(x,t) \, e^{ikx(t)-i\omega t} = \int_0^\infty dt \, <e^{ikx(t)-i\omega t}>
\end{equation}
is the Fourier representation of the propagator describing the spatial transport of particles, i.e., the Green's function of the spatial transport equation; its form depends on the (generally stochastic) behavior of $x(t)$. Generalization of the above results to three dimensions is given in Appendix A.

\section{Categorizing Acceleration Scenarios}\label{acc_scenarios}

The models constructed below are all obtained from Equation~(\ref{aa}) or equivalently from Equation~(\ref{bb}). They are grouped, in order of complexity of the equation for $P(x,t)$, as resonant, non-resonant, or resonant-broadened.

\subsection{Resonant acceleration models}\label{res_accn_models}

In resonant acceleration models, the particle trajectory is free-streaming; the stochastic equations of motion are $\dot{p}=F(x,t)$, $x=vt$.  This means that the probability function $P(x,t)$ satisfies

\begin{equation}\label{prop_resonant}
\frac{\partial P(x,t)}{\partial t}+v \, \frac{\partial P(x,t)}{\partial x}=\delta(x) \, \delta(t) \,\,\, ,
\end{equation}
with solution

\begin{equation}
P(x,t)=\delta(x-vt) \,\,\, .
\end{equation}
Equation~(\ref{aa}) then straightforwardly gives

\begin{equation}
D(p)=\int_{0}^{\infty} dt \int dx \, C(x,t) \, \delta(x-vt) = \int_{0}^{\infty}dt \, C(vt,t) \,\,\, .
\end{equation}
In this case the Lagrangian correlation function is related to the Eulerian correlation function by $C_{L}(t)=C(vt,t)$. Note that when $v \rightarrow 0$, $C_{L}(t)\rightarrow C(0,t)$, therefore the Lagrangian correlation time $\tau_L$ becomes equal to the Eulerian correlation time $\tau$, and $D(p)\rightarrow \tau <F^{2}>$.

In the Fourier representation, Equation~(\ref{prop_fourier}) shows that the propagator is

\begin{equation}\label{g1}
G(k,\omega) = \int_0^\infty dt \, e^{i(kvt-\omega t)} \,\,\, .
\end{equation}
As it stands, this integral is not convergent when $t\rightarrow \infty$. To evaluate the integral, we temporarily add an infinitesimal damping factor $\nu=0+$ in the exponential, so that

\begin{equation}\label{gg}
G(k,\omega) = \int_0^\infty dt \, e^{i(kvt-\omega t)} \, e^{-\nu t} = \frac{i}{(kv-\omega) +i\nu}
\, \buildrel {\nu \rightarrow 0} \over {\longrightarrow} \, \pi \, \delta (\omega-kv) + i {\cal P} \left ( \frac{1}{\omega-kv} \right ) \,\,\, ,
\end{equation}
where ${\cal P}$ is the principal value.  The momentum diffusion coefficient~(\ref{bb}) takes the form

\begin{equation}\label{d}
D(p)=\pi \int \int dk \, d\omega  \, S(k,\omega) \, \delta (\omega-kv) = \pi \int dk \, S(k,kv) \,\,\, .
\end{equation}

Some insight into the diffusion coefficient $D(p)$ may be earned by normalizing time and space coordinates by the Eulerian correlation time and length $\tau$ and $\lambda$, respectively. If we then define dimensionless frequency and wavenumber by $\widetilde{\omega}=\omega\tau$ and $\widetilde{k}=\lambda k$, respectively, the delta-function term in the propagator~(\ref{gg}) becomes

\begin{equation}
\pi \, \delta(\omega-kv) = \tau \, g(\widetilde{k},\widetilde{\omega};\theta) \,\,\, .
\end{equation}
Here the dimensionless propagator

\begin{equation}\label{g_definition}
g(\widetilde{k},\widetilde{\omega};\theta) = \pi \, \delta(\widetilde{\omega}-\theta\widetilde{k})
\end{equation}
is a function of the dimensionless momentum variable

\begin{equation}\label{thetadef}
\theta =\frac{v}{\lambda/\tau} \, = \frac{p}{p_0} \, ,
\end{equation}
where the reference momentum $p_0 = {m \lambda/\tau}$ has been defined. Note that $\theta$ is also the ratio of the Eulerian correlation time $\tau$ to the transport time scale $\lambda/v$.  The expression~(\ref{d}) for the diffusion coefficient becomes

\begin{equation}\label{h1}
D(p)=\left ( \frac{1}{\lambda \tau} \right ) \int \int d\widetilde{k} \, d\widetilde{\omega} \, S(\widetilde{k},\widetilde{\omega}) \, \tau \, g(\widetilde{k},\widetilde{\omega};\theta) \,\,\, .
\end{equation}
Then, noting from Equation~(\ref{normf2}) that, by definition,

\begin{equation}\label{h2}
<F^{2}> \, = \left ( \frac{1}{\lambda \tau} \right ) \int \int d\widetilde{k} \, d{\widetilde{\omega}} \, S(\widetilde{k},\widetilde{\omega}) \,\,\, ,
\end{equation}
we can eliminate the the factor $\lambda \tau$ between Equations~(\ref{h1}) and~(\ref{h2}), yielding

\begin{equation}\label{dtheta}
D(p) = \tau  \, \xi(\theta)  <F^{2}> \,\,\, ,
\end{equation}
where the dimensionless function

\begin{equation}\label{xidef}
\xi(\theta) = \frac{\int \int d\widetilde{k} \, d\widetilde{\omega} \,  S(\widetilde{k},\widetilde{\omega}) \, g(\widetilde{k},\widetilde{\omega};\theta)} {\int \int d\widetilde{k} \, d{\widetilde{\omega}} \, S(\widetilde{k},\widetilde{\omega})} \,\,\, .
\end{equation}
Since, by Equation~(\ref{lagrangian_result}), $D(p) = \tau_L <F^2>$, Equation~(\ref{dtheta}) shows that the Lagrangian correlation time is given by $\tau_{L}=\tau \, \xi (\theta)$. Hence the function $\xi(\theta)$ is simply the ratio of the Lagrangian correlation time $\tau_L$ to the Eulerian correlation time $\tau$. Further, according to Equation~(\ref{xidef}), this ratio depends only on the dimensionless momentum variable $\theta=p/p_{0}$ and on the form of the (dimensionless) force spectrum $S(\widetilde{k},\widetilde{\omega})$.

We illustrate with the Gaussian spectrum~(\ref{skwgaussian}), which, in terms of dimensionless variables, is

\begin{equation}
S(\widetilde{k},\widetilde{\omega}) = S_{0} \, e^{-\widetilde{k}^{2}/4\pi} e^{-\widetilde{\omega}^{2}/\pi} \,\,\, .
\end{equation}
Substituting this spectrum in Equation~(\ref{xidef}), we obtain

\begin{equation}
\xi(\theta) = {\frac{1} {(1+4\theta^{2})^{1/2}}} \,\,\, ;
\end{equation}
so that

\begin{equation}\label{tauelltauresdim}
D(p) =\frac{\tau \, <F^2>}{[1+4(p/p_o)^2]^{1/2}} \,\,\, .
\end{equation}
In the initial stage of the acceleration process, the particle is moving sufficiently slowly to feel only the temporal variation of the force: $\tau_{L}\sim \tau$ when $p\ll p_{0}$. If $<F^2>$ is also independent of $p$, $D(p)$ is a constant and so, by Equation~(\ref{p_growth}), $<p^{2}> \, \propto t$, as in the Obukhov model.  However, as the particle gains speed, it eventually becomes fast enough that it becomes less sensitive to the {\it temporal} variation of the force and more to its {\it spatial} variation. The Lagrangian correlation time becomes of the order of the transport time scale which is inversely proportional to the particle velocity, i.e., $\tau_L \propto p^{-1}$.  If $<F^2>$ is independent of $p$, then we can use Equation~(\ref{dprelation}) and the defining relation $\tau_{\rm acc} = p^2/D(p)$ for the acceleration time $\tau_{\rm acc}$ to obtain, in the nonrelativistic regime where the particle energy $E=p^2/2m$,

\begin{equation}\label{tauaccres}
D(p)\propto p^{-1} ; \quad \tau_{\rm acc} \propto p^3 \propto E^{3/2} \,\,\, .
\end{equation}
It should be noticed that when the force field is the gradient of a potential so that it is the potential rather than the force has a Gaussian correlation function (see Appendix A), then

\begin{equation}\label{tauaccreswave}
D(p)\propto p^{-3} ; \quad \tau_{\rm acc} \propto p^5 \propto E^{5/2} \,\,\, .
\end{equation}

\subsubsection{Resonant Acceleration by Waves}\label{res_acc_waves}

For a wave-acceleration process, we include the dispersion relation $\omega(k)$ in the Fourier spectrum of the force spectrum $S(k,\omega)$, i.e.,

\begin{equation}
S(k,\omega) = S(k) \, \delta [\omega-\omega(k)] \,\,\, .
\end{equation}
From Equation~(\ref{d}), this leads to a diffusion coefficient of the form

\begin{equation}\label{dwave}
D(p) = 2\pi \int dk \, S(k) \, \delta [\omega(k)-kv] \,\,\, ,
\end{equation}
where the factor of two takes into account waves propagating in both directions.

The nature of resonant acceleration by waves is as follows: a particle of momentum $p$ interacts with a wave if the particles's velocity is equal to the phase velocity of the wave: $v=\omega(k)/k=V_{p}$. However, stochastic resonant acceleration can occur even in the absence of a dispersion relation. Physically, this occurs because a particle still interacts with a Fourier component of the force field such that $\omega/k=v$, even though $\omega$ and $k$ may not be related by a dispersion relation $\omega(k)$. Thus, particles are stochastically accelerated provided the modes which are experienced by the particles along their trajectory have random phases. The existence of a broadband spectrum of field fluctuations, in the frame co-moving with the particles, with Fourier components not related by a dispersion relation may be identified with {\it strong turbulence}; this situation relaxes some of the restrictions imposed by a linear dispersion relation.

\citet{stu} has  shown that (a) by taking into account the dispersion relation $\omega=\omega(k)$ in the force field spectrum and (b) by enforcing conservation of the total combined energy of the field and the accelerated particles, the self-consistent quasilinear diffusion equations for wave-particle interaction are recovered \citep{sag,2003PPCF...45A.115E}. The energy gained by the resonant particles is simply lost by the waves through Landau damping. \citet{stu} uses as an example an electric force produced by Langmuir waves, but the idea readily generalizes to other forms of wave. A detailed discussion of the self-consistent quasilinear equations, with emphasize in the role of resonant and non-resonant particles, can be found in \citet{1972JPlPh...8....1K}.

In resonant acceleration by plasma waves, the momentum diffusion generally depends on the spectral properties of the turbulence. This may be seen by taking a wave frequency independent of $k$, i.e., $\omega=\omega_{0}$, and a power-law spectrum $S(k)\propto k^{-q}$ in (\ref{dwave}), then $D(p)\propto p^{1-q}$ which involves the spectral index $q$ of the turbulence\footnote{This is widely studied in the context of high-frequency turbulence of Langmuir waves, which frequency $\omega \simeq \omega _{pe}\simeq {\rm const}$ and the spectrum $S(k)\propto k^{-q}$ \citep[see, e.g.,][]{1995lnlp.book.....T}.}.  By Equation~(\ref{p_growth}), in such a scenario, the particle energy grows as $t^{2/(1 + q)}$.

Again, it is emphasized that Equation~(\ref{d}) is more general than Equation~(\ref{dwave}).  The former is valid for an arbitrary correlation function rather than one which describes only oscillations and which therefore, for a given $k$, has a peak at the frequency $\omega(k)$ of the oscillations.

\subsubsection{Remarks}\label{domain_validity}

The domain of validity of the quasilinear approximation is obtained by considering the dimensionless Kubo number $K=\tau \sqrt{<F^2>}/p_{0}$ that enters the normalized momentum equation $d{\widetilde{p}}/dt=K\widetilde{F}$, where the length, time and force have been normalized to $\tau$, $\lambda$ and $\sqrt{<F^2>}$, respectively. When $K \ll 1$, we are in the domain of validity of the quasilinear approximation.  This inequality limits the typical amplitude of the force field to

\begin{equation}
F \ll \frac{\lambda}{\tau^{2}}
\end{equation}
and gives quantitative meaning to the ``weak field limit'' discussed by \citet{stu}. Also, when the spectrum of modes is discrete, their amplitude must be sufficiently large to permit resonance overlap, and hence stochasticity, otherwise dynamical trapping of particles become substantial and acceleration is inhibited. The problem with spectral discreteness can be understood by recalling that the momentum diffusion coefficient in such a case takes the form

\begin{equation}\label{dp_discrete}
D(p)=\int_{0}^{\infty}dt\sum_{k,\omega}|F_{k,\omega}|^{2}\exp[i(kv-\omega)t] \,\,\, ,
\end{equation}
which shows that the wave-number sum inside the time integral does not decay with time, but instead exhibits recurrences \citep{2002PhR...360....1K}. These recurrences are a consequence of the periodicity of the force felt by the particles along their trajectory in the quasilinear approximation $x=vt$. This effect is similar to (but not equivalent to) the dynamical trapping of a particle oscillating in a periodic electric field where both $x(t)$ and $p(t)$ are periodic functions of time. The momentum diffusion coefficient can also be written in the form

\begin{equation}\label{dwavedis}
D(p) = \pi \sum_{k,\omega} \, |F_{k,\omega}|^{2}\, \delta
[\omega-kv] \,\,\, .
\end{equation}
This form is somewhat is pathological since diffusion occurs only over a set of separated points in velocity space.  As a result, the distribution function is not affected by the diffusion process and acceleration does not occur at all.

One way in which the singular nature of the diffusion process can be removed consists of broadening the resonant propagator $\pi \delta(\omega-kv)$ into a suitable, say Lorentzian, form:

\begin{equation}\label{dwavedis_Lorentzian}
D(p) = \sum_{k,\omega} \, |F_{k,\omega}|^{2}\, \frac{\nu}{(\omega-kv)^{2}+\nu^{2}} \,\,\, ,
\end{equation}
which may be obtained by replacing $\omega \rightarrow \omega+i\nu$ with $\nu>0$ in Equation~(\ref{dp_discrete}), see also Equation~(\ref{gg}). Another way of removing the singularity is to simply replace the discrete sum in Equation~(\ref{dp_discrete}) by a continuous one, i.e.,

\begin{equation}
\sum_{k,\omega}\rightarrow \int \frac{dk \, d\omega}{\delta k \, \delta \omega} \,\,\, .
\end{equation}
Identifying $S(k,\omega)=|F_{k,\omega}|^{2}/(\delta k \, \delta\omega )$,
one recovers Equation~(\ref{d}).

In the non-resonant or resonance-broadened acceleration models to be presented below, the propagator is no longer singular, a consequence of the seed stochasticity that is included in the description of the spatial transport of the particles and which is not, in general, dynamically related to the accelerating wave field. For this reason, stochastic acceleration occurs also when the spectrum of the wave field is singular (discrete or monochromatic) in these models.

\subsection{Non-resonant acceleration models}\label{non_res_accn_models}

Non-resonant acceleration models are based on the approximation that the spatial transport is also stochastic; see, e.g., the review by \citet{by}. In the standard case, this transport is diffusive, i.e., the particle's position varies according to a Gaussian process and the mean square displacement grows linearly with time. In other words, the stochastic equations of motion are approximated by $\dot{p}=F(x,t)$, $\dot{x}=\zeta(t)$, where $\zeta(t)$ is a Gaussian white noise with $<\zeta(t)\zeta(t')>=2\kappa \delta(t-t')$, $\kappa$ being a spatial diffusion coefficient. The propagator $P(x,t)$ satisfies the diffusion equation

\begin{equation}
\frac{\partial P(x,t)}{\partial t}-\kappa \, \frac{\partial^{2}P(x,t)}{\partial x^{2}}=\delta(x) \, \delta(t) \,\,\, ,
\end{equation}
with the standard Gaussian solution

\begin{equation}
P(x,t)=\frac{1}{\sqrt{4\pi \kappa t}} \,\, e^{-x^{2}/4\kappa t} \,\,\, .
\end{equation}
Particles perform a standard random walk, with $<x^2> = 2\kappa t$; the characteristic diffusion time that is associated with the correlation length $\lambda$ is $\tau_D = \lambda^2/\kappa$. In Fourier space, the propagator

\begin{equation}
G(k,\omega) = \int_0^\infty dt \, <e^{i[kx(t)-\omega t]}> = \int dt \, e^{-i \omega t} e^{-k^{2}<x^{2}>/2} = \frac{1}{i\omega +\nu_{d}} = {\frac {\nu_d - i \omega} {\omega^2 + \nu_d^2}} \,\,\, ,
\end{equation}
where the damping factor $\nu_{d}=\kappa k^{2}$, a parameter associated with the spatial diffusive transport.  For a force field with a continuous spectrum of Fourier components, the diffusion coefficient becomes

\begin{equation}\label{trans_broad}
D(p)=\int \int dk \, d\omega \, S(k,\omega) \, \frac{\kappa k^{2}}{\omega^{2}+(\kappa k^{2})^{2}} \,\,\, ,
\end{equation}
and involves a Lorentzian of width $\nu_{d}^{-1} = (\kappa k^{2})^{-1}$. A similar expression exists for force fields that consist of discrete sum of Fourier components:

\begin{equation}\label{trans_broad_discrete}
D(p)=\sum_k \sum_\omega \, \vert \widehat{F}_{k,\omega} \vert^2 \, \frac{\kappa k^{2}}{\omega^{2}+(\kappa k^{2})^{2}} \,\,\, .
\end{equation}
If we again normalize time and space by the Eulerian correlation scales $\tau$ and $\lambda$, then

\begin{equation}
\frac{\kappa k^{2}}{\omega^{2}+(\kappa k^{2})^{2}} = \tau \, g(\widetilde{k},\widetilde{\omega};\zeta) \,\,\, ,
\end{equation}
where the dimensionless propagator

\begin{equation}
g(\widetilde{k},\widetilde{\omega};\zeta) = \frac{\zeta \widetilde{k}^{2}}{\widetilde{\omega}^{2}+(\zeta\widetilde{k}^{2})^{2}}
\end{equation}
is a function of the dimensionless parameter

\begin{equation}\label{zetadef}
\zeta = \frac{\tau}{\lambda^{2}/\kappa}=\frac{\tau}{\tau_D} \,\,\, ,
\end{equation}
the ratio of the Eulerian and transport time scales. Again, we can write the momentum diffusion coefficient in the form

\begin{equation}
D(\zeta)= \tau \, \xi(\zeta)<F^{2}> \,\,\, ,
\end{equation}
where the non-dimensional parameter

\begin{equation}
\xi(\zeta)= \frac{\int \int d\widetilde{k} \, d\widetilde{\omega} \, S(\widetilde{k},\widetilde{\omega}) \, g(\widetilde{\omega},\widetilde{k};\zeta)} {\int \int d\widetilde{k} \, d\widetilde{\omega} \, S(\widetilde{k},\widetilde{\omega})} \,\,\, .
\end{equation}
When $\zeta\ll 1$, i.e., $\tau\ll \tau_{D}$, then the propagator reduces to the delta function, $g(\widetilde{k},\widetilde{\omega};\zeta\ll 1)\rightarrow \pi \, \delta(\widetilde{\omega})$, as in the Obukhov model. On the other hand, when $\zeta\gg 1$, i.e., $\tau \gg \tau_{D}$, the dimensionless propagator becomes $g(\widetilde{k},\widetilde{\omega};\zeta\gg1) \rightarrow (\zeta \widetilde{k}^{2})^{-1}$. Exact forms for $\xi(\zeta)$ may be obtained for a prescribed spectrum $S(\widetilde{k},\widetilde{\omega})$ of the force field. However, it is sufficient to note that when $\zeta\ll 1$, $\xi(\zeta)\propto \zeta^{0}$, and that when $\zeta \gg 1$, $\xi(\zeta)\propto \zeta^{-1}$. Thus, the results are summarized as follows: in the weak diffusion limit ($\tau \ll \tau_{D}$),

\begin{equation}
D(p) \sim \, \tau <F^2> \,\,\, ,
\end{equation}
while, in the strong diffusion limit ($\tau \gg \tau_{D}$),

\begin{equation}
D(p) \sim  \tau_{D} <F^2> \,\,\, .
\end{equation}
As for resonant acceleration, the Lagrangian correlation time interpolates between the Eulerian and the transport time scale, the latter being now the diffusive time scale. Therefore, if the force field and the spatial diffusion coefficient $\kappa$ (and hence the transport time scale $\tau_D$) are independent of momentum, in both weak and strong diffusion limits $D(p)$ is independent of $p$.

To summarize, for non-resonant acceleration, in the nonrelativistic limit,

\begin{equation}\label{tauaccnonres}
D(p)\propto p^{0} ; \quad \tau_{\rm acc} \propto p^2 \propto E \,\,\, .
\end{equation}

\subsubsection{Fractional-Order Fermi acceleration}\label{frac_order_fermi}

Anomalous momentum gain with $< p^{2}>$ growing faster than linearly in time can occur in non-resonant models, even when the force field and the transport time scale are independent of momentum. This may happen for a a static ($\partial/\partial t = 0$) force field, i.e., $S(k,\omega)=S(k) \, \delta (\omega)$, for which by Equation~(\ref{trans_broad}),

\begin{equation}
D(p)= {\frac{1}{\kappa}} \, \int \frac{ S(k) \, dk}{k^{2}} \,\,\, .
\end{equation}
Moreover, let us take the force field to be scale-free, for instance

\begin{equation}
S(k)=S_0 = {\rm constant} \,\,\, ,
\end{equation}
so that the integral for $D(p)$ diverges at small $k$ like $k^{-1}$
and $D(p) \rightarrow \infty$. It follows that for a scale-free force, the momentum gain process is super-diffusive:

\begin{equation}
<p^{2}> \, \propto t^{\beta} \,\,\, ; \qquad \beta >1 \,\,
\end{equation}
at all times.  However, if the force has a characteristic scale $\lambda$, the super-diffusive process is limited to times less than the time scale $\tau_D = \lambda^{2}/\kappa$ taken for spatial diffusion over a length $\lambda$ to occur.  Matching the super-diffusive momentum gain regime $<p^2> \, \sim t^\beta$ to its diffusive counterpart $<p^2> \, \sim \lambda t \sim \sqrt{\kappa \tau_D} \, t$ at $t=\tau_D$, we see that $\tau_D^\beta \sim \tau_D^{1/2} \tau_D$, so that $\beta = 3/2$.  Hence

\begin{equation}
<p^{2}> \sim t^{3/2} \,\,\, ; \quad t<\tau_D; \qquad <p^2> \sim t \,\,\, ; \quad t>\tau_D \,\,\, .
\end{equation}
Summarizing,

\begin{equation}\label{tauaccfracfermi}
D(p) \rightarrow \infty \,\,\, ;  \quad \tau_{\rm acc} \propto p^{4/3} \propto E^{2/3} \,\,\, .
\end{equation}

Superdiffusive momentum transport occurs in scale-free force fields and is the preasymptotic regime of non-resonant acceleration models involving frozen fields. The reason is that the random motion of the particles results in multiple returns to any given position.  Because of the static nature of the force field, the particle experiences the same value of the accelerating field at each visit to the same position. Such recurrences introduce correlations in the stochastic acceleration process that are absent in a non-static environment, rendering the situation more akin to a systematic one, and hence faster than diffusive. We may thus call such a stochastic acceleration process {\it Fermi acceleration of fractional order}. The essential difference between second-order Fermi acceleration and  fractional-order Fermi acceleration is as follows.  For the second-order process $D(p)$ exists.  It may produce anomalous (superdiffusive) behavior because of the dependence of $D(p)$ on $p$ [e.g., $D(p) \propto p; <p^{2}>\propto t^{2}$; see Equations~(\ref{dprelation}) and (\ref{p_growth})]; however, the evolution of $f(p,t)$ is still given by a standard Fokker-Planck equation. However, in fractional-order Fermi acceleration, the anomalous superdiffusive behavior in momentum space is due to the fact that the momentum diffusion coefficient is infinite, and the the equation governing the evolution of $f(p,t)$ is not the Fokker-Planck equation.  Stochastic acceleration models based on the fractional Fokker-Planck equation have been discussed by \citet{bi}.

\subsubsection{Nonresonant Stochastic Acceleration by Turbulent Waves}\label{nonres_accn_waves}

When $\omega$ and $k$ obey a dispersion relation, it follows from writing $S(k,\omega) = S(k) \, \delta [\omega-\omega(k)]$ that the momentum diffusion coefficient, when it exists, is

\begin{equation}
D(p) = \int dk \, S(k) \, \frac{\kappa k^2}{\omega^2(k)+(\kappa k^2)^2} \,\,\, .
\end{equation}
Taking the simple dispersion relation $\omega(k) = V_p k$, and defining the dimensionless parameter

\begin{equation}
\zeta_W = \frac {\lambda V_p} {\kappa} \, = {\frac{\tau_D} { \tau_{W}}} \,\,\, ,
\end{equation}
with $\tau_{W}=\lambda/V_p$, yields the following form for the momentum diffusion coefficient:

\begin{equation}
D(\zeta_W) = \tau_D \, \xi(\zeta_W) \,<F^2> \,\,\, .
\end{equation}
Here the dimensionless function

\begin{equation}
\xi(\zeta_W) = \frac {\int d\widetilde{k} \,  S(\widetilde{k}) \, g(\widetilde{k} ; \zeta_W)} {\int d\widetilde{k} \, S(\widetilde{k}) }
\end{equation}
and

\begin{equation}
g(\widetilde{k};\zeta_W)= \frac{\widetilde{k}^{2}}{\zeta_W^{2}\widetilde{k}^{2}+\widetilde{k}^{4}} \, \,\, .
\end{equation}
Again, an exact expression for $\xi(\zeta_W)$ may be obtained for a specified form of the wavenumber spectrum $S(\widetilde{k})$. In the limit $\zeta_W \ll 1$, $g(\widetilde{k},\zeta_W \ll 1) \rightarrow \widetilde{k}^{-2} \, ,$ leading to $ \xi(\zeta_W \ll 1) \rightarrow \, {\rm constant}$. In the limit $\zeta_W \gg 1$, $g(\widetilde{k},\zeta_W \gg 1)\rightarrow \zeta_W^{-2}$ and thus $ \xi(\zeta_W \gg 1) \rightarrow  \, \zeta_W^{-2} \,$. These results are summarized as follows. When $\tau_{D} \ll \tau_{W}$, then

\begin{equation}
D(p) \sim \, \tau_{D} <F^2> \ ,
\end{equation}
while, in limit $\tau_{D} \gg \tau_{W}$,

\begin{equation}
D(p) \sim  \frac{\tau_{W}^{2}}{\tau_{D}} <F^2>.
\end{equation}
In both cases, therefore,

\begin{equation}\label{tauaccnonreswave}
D(p)\propto p^{0} ;  \quad \tau_{\rm acc} \propto p^2 \propto E \,\,\, .
\end{equation}

\subsubsection{Nonresonant stochastic acceleration by a monochromatic wave field}

We have already given -- Equation~(\ref{trans_broad_discrete}) -- a general expression for the momentum diffusion coefficient which results from \emph{non-resonant} acceleration by any periodic force field consisting of a discrete spectrum of Fourier components. Here we discuss the illustrative example of a monochromatic mode with frequency $\omega_{0}$ and wave-number $k_{0}$. From Equation~(\ref{trans_broad_discrete}), the momentum diffusion coefficient is given in this case by

\begin{equation}\label{trans_monoprice}
D(p)=\vert \widehat{F}_{k_{0},\omega_{0}} \vert^2 \, \frac{\kappa k_{0}^{2}}{\omega_{0}^{2}+(\kappa k_{0}^{2})^{2}} \,\,\,
\end{equation}
and the frequency can be taken to obey the linear dispersion relation $\omega_{0}=V_{p}k_{0}$. This result provides a simple explanation of the reason why particles are stochastically accelerated by adiabatic compressive forces $F=-\frac{1}{3}p \, \nabla \cdot \mathbf{V}$ (see Section~\ref{parker}), even if the flow $\mathbf{V}$ consists of smooth and regular compressions and expansions. This mechanism was called {\it diffusive compression acceleration} by \citet{2003ICRC....6.3685J}, where it was noted that the acceleration mechanism has similarities with both second-order Fermi and diffusive shock accelerations; see also \citet{2010JGRA..11512102Z}. Therefore, a spatially periodic compressive flow produces the same kind of stochastic acceleration as a strongly turbulent flow or a shock flow. A legitimate question is thus: what is the form of the Lagrangian correlation function, or equivalently, what is the shape of the frequency spectrum of the monochromatic wave field in the frame of the random walking particle? A simple calculation shows that this spectrum is broad and has the Lorentzian form:

\begin{equation} \label{osc}
I(\omega)=\frac{\vert \widehat{F}_{k_{0},\omega_{0}}\vert ^{2}}{\pi} \,\frac{\kappa k_{0}^{2}}{(\omega-\omega_{0})^{2}+(\kappa
k_{0}^{2})^{2}} \,\,\, .
\end{equation}
Note that this spectrum is similar to the one of an harmonic oscillator in which the frequency is randomly perturbed in time, i.e., the Anderson-Kubo oscillator \citep{1963JMP.....4..174K}:

\begin{equation}
\dot{\psi}(t)=i\omega(t)\psi\,\, ; \qquad  \omega(t)=\omega_{0}+\Delta \omega(t) \,\,\, .
\end{equation}
The frequency perturbations are assumed to have a zero average and a correlation function given by

\begin{equation}
<\Delta \omega(t) \, \Delta
\omega(t')> \, =\frac{D_{\omega}}{\tau}\exp(-|t-t'|/\tau) \,\,\, .
\end{equation}
In the white noise limit $\tau\rightarrow 0$, the spectrum is indeed a Lorentzian:

\begin{equation}
I(\omega)=\frac{1}{\pi}
\,\frac{D_{\omega}}{(\omega-\omega_{0})^{2}+D_{\omega}^{2}} \,\,\, ;
\end{equation}
however, in the limit of large $\tau$, i.e., $(D_{\omega}\tau)^{1/2}\gg 1$, we obtain the Gaussian spectrum

\begin{equation}
I(\omega)=
\left( \frac{2\pi D_{\omega}}{\tau}\right)^{-1/2} \, \exp\left[\frac{-(\omega-\omega_{0})^{2}}{2D_{\omega}/\tau}\right] \,\,\, ,
\end{equation}
showing how the broadening of the line shape depends on the correlation time of the frequency perturbations.

\subsubsection{Remarks}\label{res_vs_nonres}

In its original version, the quasilinear diffusion equation is a Fokker-Planck equation that describes the diffusion of particles in momentum space due to the action of turbulent waves. The quasilinear formalism can be seen in a general perspective by casting it in Hamiltonian form. In absence of the the accelerating field, the particle motion is described by an unperturbed Hamiltonian $H_{0}$ which is independent of time. This means that the unperturbed motion occurs at constant energy and is rectilinear in an infinite homogeneous plasma. The total single-particle Hamiltonian is then $H_{0}+H_{1}(t)$ and the change in the particle's energy is a consequence of the time-dependent perturbation $H_{1}(t)$ with $H_{1}/H_{0}\sim \epsilon \ll 1$. In standard quasilinear theory, the stochasticity comes from $H_{1}(t)$, i.e., from the assumption of random phases in the broad spectrum of the accelerating field. The smallness parameter $\epsilon$ expresses the condition that the amplitude of the force is small, so that the acceleration efficiency can be derived using linear $O(\epsilon)$ theory, hence the terminology ``quasilinear.'' It is not necessary to assume an infinite, homogeneous plasma, the fundamental requirement is simply that $H_{0}$ is integrable. Quasilinear diffusion occurs in action space; the unperturbed motion in angle space is also rectilinear, just as that for the motion in coordinate space in a homogeneous medium. The accelerating waves are in resonance with the periodic bounce motion of the particles in this case

However, there is also another form of quasilinear theory which can be applied in the complete reverse case to that above, namely when the unperturbed Hamiltonian is {\it non}-integrable, giving rise to \emph{chaotic} motion of the particles in the absence of the accelerating field. The unperturbed dynamics described by $H_{0}$ provides the source of randomization for stochastic acceleration to occur under the action of the time-dependent perturbation $H_{1}$. Notably, since the unperturbed motion is a seed stochasticity, stochastic acceleration can occur also when the perturbation is coherent. Here, we have labeled such quasilinear models as ``resonant'' or ``non-resonant.'' The essence of non-resonant stochastic acceleration is therefore simply understood by considering the unperturbed motion to be a spatial random walk, with a mean free path of the particles smaller than the correlation length of the field.

\subsection{Resonance-broadened models}\label{res_broadened}

In resonance-broadened models, the spatial transport is a combination of free-streaming and spatial dispersion: $x=vt+\Delta x$, where $\Delta x$ is a random walk. The idea of resonance broadening in the context of acceleration
of particles was introduced by \citet{dup} [see also \citet{wein,rud,benf}].  The principal effect of the dispersion can be seen by applying an orbit diffusion $<\Delta x^{2}>=2\kappa t$ around the free-streaming solution; hence, the propagator $P(x,t)$ evolves according to

\begin{equation}\label{fppp1}
\frac{\partial P(x,v,t)}{\partial t}+v \, \frac{\partial P(x,v,t)}{\partial x}-\kappa \, \frac{\partial^{2}P(x,t)}{\partial x^{2}}=\delta(x) \, \delta(t) \,\,\, .
\end{equation}
The solution for the propagator $P(x,t)$ is

\begin{equation}\label{advdiff}
P(x,t)={\frac{1} {\sqrt{4\pi \kappa t}}} \, \exp\left(-\frac{(x-vt)^{2}}{4\kappa t}\right) \,\,\, .
\end{equation}
When $\kappa=0$, $P(x,v,t)$ reduces to $\delta(x-vt)$, which is the usual resonance function of quasilinear theory, considered in Section~\ref{res_accn_models}.  Hence, the essential effect of the diffusive term is to produce a broadening of the resonance function.

Equivalently, the expression

\begin{equation}
D(p)=\int \int dk \, d\omega \int_{0}^{\infty} dt \, S(k,\omega) \, e^{-i\omega t} <e^{ikx(t)}> \,\,\, ,
\end{equation}
is calculated with

\begin{equation}
x(t)=vt+\Delta x(t) \,\,\, ,
\end{equation}
where $\Delta x(t)$ is the deviation from free-streaming motion, which is assumed  to be a standard random walk with $<\Delta x^2> = 2\kappa t$. The momentum diffusion coefficient now involves a velocity-shifted Lorentzian:

\begin{equation}
D(p)=\int \int dk \, d\omega \, S(k,\omega)\, \frac{\kappa k^{2}}{(\omega-kv)^{2}+(\kappa k^{2})^2} \, \,\, .
\end{equation}
Scaling as usual, we may write the momentum diffusion coefficient in the form

\begin{equation}
D(\theta;\zeta)= \tau \, \xi(\theta; \zeta) <F^{2}> \,\,\, ,
\end{equation}
where $\theta$ and $\zeta$ are as defined in Equations~(\ref{thetadef}) and~(\ref{zetadef}), respectively:

\begin{equation}
\xi(\theta;\zeta)= \frac{\int \int d\widetilde{k} \, d\widetilde{\omega} \, S(\widetilde{k},\widetilde{\omega}) \, g(\widetilde{\omega},\widetilde{k};\theta;\zeta)} {\int \int
d\widetilde{k} \, d{\widetilde{\omega}} \, S(\widetilde{k},\widetilde{\omega})} \,\,\, ,
\end{equation}
and

\begin{equation}
g(\widetilde{k},\widetilde{\omega};\theta;\zeta)= \frac{\zeta \widetilde{k}^{2}}
{(\widetilde{\omega}-\theta \widetilde{k})^{2}+(\zeta \widetilde{k}^{2})^{2}} \,\,\, .
\end{equation}
This result now involves {\it two} non-dimensional parameters $\theta$ and $\zeta$:  the model has both a resonant and non-resonant character depending on the relative value of these parameters. Obviously, when both are small we recover the Obukhov model.

\subsubsection{Dupree theory}\label{dupree_theory}

In formal resonant-broadening theories the resonance is broadened by a function $R(\omega-kv,\nu)$, not necessarily a Lorentzian, which tends to $\delta(\omega-kv)$ when the frequency $\nu\rightarrow 0$. In the theory of \citet{dup}, the accelerating field is idealized by position-independent Gaussian white noise, i.e., the Obukov model, leading to a spatial dispersion with $<\Delta x^{2}>=(2/3)Dt^{3}/m^2$ around the free-streaming motion. Therefore, the momentum diffusion coefficient originally derived by \citet{dup} is

\begin{equation}\label{dup}
D(p)=\int dk \, d\omega \int_{0}^{\infty} dt \, S(k,\omega) \, e^{i(kvt-\omega t)} \, e^{-k^{2}Dt^{3}/3 m^2} \,\,\, .
\end{equation}
Here, the spatial transport in the accelerating field is dynamically correlated with the field itself. It is important to notice that a characteristic result of this kind of analysis is an implicit, rather than explicit, relation for $D$, which appears on both sides of Equation~(\ref{dup}). Formally, this means that in order to obtain the momentum diffusion coefficient $D$, we need to solve an equation of the form

\begin{equation}
D(p)=\tau_{L}(D) <F^2> \,\,\, ,
\end{equation}
with $\tau_{L}$ a function of $D$; thus $D$ does {\it not} straightforwardly scale as the square of the amplitude of the force.

\subsubsection{Remarks}\label{rem1}

To conclude this section, we mention another form of broadening. In the presence of wave turbulence, for a given wave-number the frequency spectrum does not have to be sharply peaked at the frequency corresponding to the dispersion relation $\omega=\omega(k)$. When the spectrum of the force field has a peak around $\omega(k)$ with width $\tau_{c}^{-1}$, the wave spectrum can be modeled as a broadened delta-function, for instance \citep{by} as a Lorentzian

\begin{equation}\label{turb_broad}
\mid {\hat F}(k,\omega)\mid^{2} = S(k,\omega) \, \delta [\omega-\omega(k)]\rightarrow \pi^{-1} \, S(k) \, \frac{\tau_{c}^{-1}}{[\omega-\omega(k)]^{2}+(\tau_{c}^{-1})^{2}} \,\,\, .
\end{equation}
Broadening of the propagator $\delta(\omega-kv)$ and broadening of the frequency spectrum $\delta[\omega-\omega(k)]$ are formally similar. However, since they involve two different processes they may still be assigned two distinctly different physical meanings. The first is a result of the spatial transport of particles; the second is a spectral property of the turbulent field involving for instance non-linear interactions between the waves themselves. In the context of particle transport, various mechanisms of resonance broadening, including wave damping or dynamical turbulence have been discussed by \citet{1993JPlPh..49...63S} and \citet{1994ApJ...420..294B}.

\subsection{Non-Gaussian spatial transport}\label{non_gaussian}

Up to now we have assumed that the stochastic part of the spatial transport was described by a standard diffusion equation. {However spatial transport can described by a variety of models, including the Chapman-Kolmogorov and the telegrapher equations \citep{2006ApJ...651..211W}.} Let us take, for instance, the propagator to be the solution of the fractional diffusion equation

\begin{equation}
D^{\beta}_{t} \, P(x,t) = \kappa \, D^{\Gamma}_{\mid x\mid} \, P(x,t) \,\,\, .
\end{equation}
This produces anomalous dispersion\footnote{The standard Gaussian case is recovered with $\beta=1$ and $\Gamma=2$.} with

\begin{equation}
<x^{2}> \, \propto \, t^{2\beta/\Gamma}
\end{equation}
and a corresponding non-Gaussian probability distribution functions $P(x,t)$. For instance, resonance broadening by super-diffusive transport (Levy flights) can be modeled by the following advection-diffusion equation:

\begin{equation}
\frac{\partial P(x,t)}{\partial t} + v \, \frac{\partial P(x,t)}{\partial x}=\kappa \, D^{\Gamma}_{\mid x\mid} \, P(x,t) \,\,\, .
\end{equation}
Since the Fourier transform is ${\cal F}(D^{\Gamma}_{\mid x\mid}f(x))=\mid k\mid^{\Gamma} f(k)$, we obtain

\begin{equation}
D(p)=\int \int dk \, d\omega\, S(k,\omega) \, \frac{\kappa |k|^{\Gamma}}{[\omega(k)-kv]^{2}+(\kappa |k|^{\Gamma})^{2}} \,\,\, .
\end{equation}
The corresponding transport time scale is correspondingly different from the diffusive timescale $\nu_{d}^{-1}$ found earlier.

\subsection{Anisotropic spatial transport}\label{anisotropic_transport}

We now discuss a simple variation of the cases discussed above. We still assume the existence of a strong guide field $\mathbf{B}_{0}$ directed along the $x$-axis and the existence of a stochastic force parallel to the magnetic field $F(x,y,z,t)$, say a parallel electric force or magnetic mirror force. The particle transport in this field is taken to be mainly free-streaming {\it along} the magnetic field but diffusive {\it across} the field. Hence, the propagator $P(x,y,z,t)$ satisfies the advection-diffusion equation

\begin{equation}
\frac{\partial P(x,r_\perp,t)}{\partial t}+v_{\parallel}\frac{\partial P(x,r_\perp,t)}{\partial x}=\kappa_{\perp} \nabla_\perp^2 P(x,r_\perp,t)\,\,\, ,
\end{equation}
where $\nabla_\perp$ is the two-dimensional Laplacian in the directions perpendicular to the streaming. The solution is similar to Equation~(\ref{advdiff}), with $\kappa_{\perp}$ being the cross-field spatial diffusion coefficient. The corresponding momentum diffusion coefficient is given by

\begin{equation}
D(p_{\parallel}) = 2 \pi \int k_{\perp} \, dk_{\perp} \int dk_{\parallel}\int d\omega \, S(k_{\perp},k_{\parallel},\omega) \, \frac{ \kappa_{\perp} k_{\perp}^{2}} {(\omega-k_{\parallel}v_{\parallel})^{2}+(\kappa_{\perp} k_{\perp}^{2})^{2}} \,\,\, ,
\end{equation}
where the turbulence is also assumed to be isotropic in the plane perpendicular to $\mathbf{B}_{0}$. As expected, perpendicular diffusion alone can act to broaden the resonance between the modes and the particles. Moreover, if these modes are waves with frequency $\omega$ which is related to the wavenumber $\bf{k}$ through the anisotropic dispersion relation $\omega=\omega(k_{\perp},k_{\parallel})$, then the force field spectrum $S(k_{\perp},k_{\parallel},\omega)=S(k_{\perp},k_{\parallel})\delta [\omega-\omega(k_{\perp},k_{\parallel})]$, and

\begin{equation}
D(p_{\parallel}) = 2\pi \int k_{\perp} \, dk_{\perp} \int dk_{\parallel} \, S(k_{\perp},k_{\parallel}) \, \frac{ \kappa_{\perp} k_{\perp}^{2}} {[\omega(k_{\perp},k_{\parallel})-k_{\parallel}v_{\parallel}]^{2}+(\kappa_{\perp} k_{\perp}^{2})^{2}} \,\,\, .
\end{equation}

\section{Forms of the Accelerating Force Field}\label{forms}

We now move on to consider various possible type of force fields and the resulting forms of $D(p)$. Recall from Equation~(\ref{lagrangian_result}) that in general $D(p)=\tau_L \, <F^2>$, and it should be noted that the dependence of $D$ on $p$ can arise both through the Lagrangian correlation timescale $\tau_L$ and/or because $<F^2>$ is itself momentum-dependent.

\subsection{Drift-kinetic equations -- magnetic mirror and parallel electric forces}\label{drift_kinetic}

Since turbulent energy is generally a decreasing function of frequency in magnetized plasmas, it is reasonable to assume that low-frequency turbulence is most efficient at accelerating particles. For electromagnetic fluctuations that satisfy $\omega\ll \omega_{ci}$, where $\omega_{ci}$ is the ion gyrofrequency, the motion of particle gyrocenters obeys the drift kinetic equations

\begin{equation}
\frac{d
\mathbf{X}}{dt}=v_{\parallel}\mathbf{b}+\mathbf{v}_{\mathbf{E} \times \mathbf{B}} \,\,\, ,
\end{equation}
\begin{equation}
\frac{d v_{\parallel}}{dt}=\frac{q}{m}E_{\parallel}-\mu_M \nabla_{\parallel}B \,\,\, ,
\end{equation}
\begin{equation}
\frac{d \mu_M}{dt}=0 \,\,\, .
\end{equation}
Here $\mathbf{v}_{\mathbf{E} \times \mathbf{B}}=c(\mathbf{E}\times \mathbf{B})/B^2$ is the $\mathbf{E} \times \mathbf{B}$ drift velocity, $\mathbf{b}=\mathbf{B}/B$, $B=\mid\mathbf{B}\mid$ and the magnetic moment

\begin{equation}
\mu_M=\frac{v_{\perp}^{2}}{2B} \,\,\, .
\end{equation}
The corresponding kinetic equation is thus

\begin{equation}
\frac{\partial f}{\partial t}+ \left (v_{\parallel}\mathbf{b}+\mathbf{v}_{\mathbf{E} \times \mathbf{B}} \right ).\nabla f+ \left (\frac{q}{m}E_{\parallel}-\mu_M\nabla_{\parallel}B \right )\frac{\partial f}{\partial v_{\parallel}}=0 \,\,\, .
\end{equation}
This shows that low-frequency turbulence with $\omega\ll \omega_{ci}$ accelerates particles both through the fluctuating parallel electric force

\begin{equation}
F_{\parallel}=qE_{\parallel}
\end{equation}
and the fluctuating magnetic-mirror force

\begin{equation}\label{mir}
F_{\parallel}=\frac{m v^{2}_{\perp}}{2B} \, \nabla_{\parallel} B \,\,\, .
\end{equation}
These two parallel forces are responsible for Landau damping (or transit-time damping) of various plasma waves via resonant acceleration of the particles. In general, however, particle acceleration through these stochastic forces can be either resonant or non-resonant. In the latter case, zero-frequency modes are absorbed through particle acceleration by a process akin to Ohmic or viscous damping. Acceleration by these forces is also intrinsically anisotropic. To produce an isotropic distribution function, therefore, a high level of pitch-angle scattering may also be required. Thus, a scenario can be envisaged where one component of the turbulent fluctuations produces, on a fast time-scale, the isotropization of the distribution function via pitch-angle scattering. This results in a diffusive transport of the particles at constant energy with a spatial transport coefficient which is in general a function of the particle speed. Then, another component of the turbulence acts on this diffusive motion to produce the energy change, resulting in stochastic acceleration of the particles. This is essentially the \citet{parker} formalism, where energy changes are a result of large scale compressions.

\subsection{Finite Larmor-radius effects}\label{flr}

In the drift-kinetic approximation, the electromagnetic field fluctuations are an average of the true fluctuations over the fast time scales associated with the gyromotion of the particles. Therefore, in this approximation, the particles feel an effective low-frequency field which acts on their gyrocenter. For $k_{\perp}\rho \ll 1$ (where $\rho=v_{\perp}/\omega_{c}$ is the gyroradius), the particle position and the gyrocenter position are essentially the same. However, for $k_{\perp}\rho\geq 1$, the effect of the finite gyroradius must be taken into account. For circular motion, the gyrocenter position $\mathbf{X}$ is related to the particle position $\mathbf{x}$ via

\begin{equation}
\mathbf{X}=\mathbf{x}-\boldsymbol\rho(\varphi) \,\,\, ,
\end{equation}
where $\varphi$ is the phase and $\boldsymbol\rho$ is the gyroradius vector defined by

\begin{equation}
\boldsymbol\rho=\frac{\mathbf{b}_{0}\times \mathbf{v}_{\perp}}{\omega_{c}} \,\,\, .
\end{equation}
Upon introducing the Fourier representation $F(\mathbf{x},t) = \int \int
d\mathbf{k} \, d\omega \, {\hat F}(\mathbf{k},\omega) \, e^{i(\mathbf{k} \cdot \mathbf{x}-\omega t)}$ of an arbitrary field $F(\mathbf{x},t)$, one finds that

\begin{equation}
<F(\mathbf{x},t)>_{\varphi} \, \equiv \overline{F}(\mathbf{X},t)=\int d\mathbf{k} \, d\omega \, {\hat F}(\mathbf{k},\omega) \, e^{i(\mathbf{k} \cdot \mathbf{X}-\omega t)} \, \frac{1}{2\pi}\int d\varphi \, e^{i\mathbf{k} \cdot \boldsymbol\rho(\varphi)} \,\,\, .
\end{equation}
Since $\mathbf{k} \cdot \boldsymbol\rho(\varphi)=k_{\perp}\rho\cos \varphi$, one can make use of the generating function for Bessel functions $e^{iz \cos\varphi}=\Sigma_{n=-\infty}^{n+\infty} \, i^{n} \, J_{n}(z) \, e^{in\varphi}$ to obtain

\begin{equation}
<F(\mathbf{x},t)>_{\varphi} \, =\int d\mathbf{k} \, d\omega \, J_{0}(k_{\perp}\rho) \, {\hat F}(\mathbf{k},\omega) \, e^{i(\mathbf{k} \cdot \mathbf{X}-\omega t)} \,\,\, .
\end{equation}
We see that the effective field felt by the particle is reduced by a factor $J_{0}(k_\perp\rho)$ which depends on the perpendicular wave-number; this correction is a finite-Larmor-radius (FLR) effect. The FLR effect is most simply accounted for in the Fourier representation where it enters as a simple renormalization of the force field spectrum

\begin{equation}
S(\mathbf{k},\omega)\rightarrow J^{2}_{0}(k_{\perp}\rho) \, S(\mathbf{k},\omega) \,\,\, .
\end{equation}
Moreover, we can define the {\it gyrocenter propagator} as

\begin{equation}
G(\mathbf{k},\omega) =\int_0^\infty dt \,
<e^{i\mathbf{k} \cdot \mathbf{X}(t)-i\omega t}> \,\,\, .
\end{equation}
The FLR modified momentum diffusion coefficient takes the form

\begin{equation}
D= \int \int d\mathbf{k} \, d\omega \, J^{2}_{0}(k_{\perp}\rho) \, S(\mathbf{k},\omega) \, G(\mathbf{k},\omega) \,\,\, .
\end{equation}
In the case where the unperturbed guiding-center motion is rectilinear at constant speed $v_{\parallel}$ along the magnetic field, $\mathbf{k} \cdot \mathbf{X}(t)=k_{\parallel}v_{\parallel}t$, and therefore

\begin{equation}\label{dgr}
D(p_{\parallel},p_{\perp}) = \pi \int \int d\mathbf{k} \, d\omega \, J^{2}_{0} \left ( \frac{k_{\perp}v_{\perp}}{\omega_{c}} \right ) \, S(\mathbf{k},\omega) \, \delta(\omega-k_{\parallel}v_{\parallel}) \,\,\, .
\end{equation}

The momentum diffusion coefficient (\ref{dgr}) involved in the Landau resonance between the field and the particles is written here in terms of the parallel and perpendicular momentum components $p_{\parallel}$ and $p_{\perp}$. Alternatively, it can be expressed in terms of the two independent variables $p$ and $\mu$, where $p=\sqrt{p_{\perp}^{2}+p_{\parallel}^{2}}$ is the magnitude of the momentum and $\mu=p_{\parallel}/p$ is the direction pitch-angle cosine. For computational applications, it can also be useful to replace the Bessel function by its Pad\'e approximation as \citep[see, e.g.,][]{1992nrca.book.....P}:

\begin{equation}
J_{0}(k_\perp\rho)\sim \frac{1}{1+(k_{\perp}\rho)^{2}/4} \,\,\, .
\end{equation}

It is important to include the FLR renormalization of the field spectrum when describing the acceleration and transport of particles by low-frequency turbulence in situations where the field fluctuations have significant energy at perpendicular wave numbers, i.e., at wavelengths of the order of or larger than $\rho$. \citet{2010PhPl...17f2308B} have discussed the spectral structure of the parallel electric field associated with strong anisotropic Alfv\'enic turbulence \citep{1995ApJ...438..763G}, showing that the latter attains maximum energy when $k_{\perp}\rho_{i}\sim 1$. The frequency of such anisotropic fluctuations remains well below the ion cyclotron frequency even at $k_{\perp}\rho_{i}\sim 1$ \citep{2006ApJ...651..590H,2009ApJS..182..310S}. At the cyclotron frequency, another FLR effect comes into play which is the interaction between the gyromotion of the particles and perpendicular electric fields. The full expression for the quasilinear momentum diffusion tensor, including both the Landau resonance and the gyroresonances can be found in \citet{1966PhFl....9.2377K}, \citet{1992JPhG...18.1089J} and \citet{1993JPlPh..49...63S}. New forms of the gyrophase averaged Fokker-Planck coefficients for particle transport have recently been discussed by \citet{2011ApJ...732...96S} and \citet{2012ApJ...745..153C}.

\subsection{Acceleration by magnetic mirrors produced by fast-mode waves}\label{accn_by_mirrors}

A popular model for flares is the acceleration by the fluctuating magnetic mirror force associated with fast mode turbulence \citep{kul,acht,mila,miller,si,yan}. For resonant acceleration by waves which possess a parallel electric or magnetic mirror force (i.e., waves that are subject to Landau or transit-time damping), we have

\begin{equation}\label{an}
D(p_{\parallel},p_{\perp}) = 4\pi^{2}\int k_{\perp} \, dk_{\perp}\int dk_{\parallel} \, S(k_{\perp},k_{\parallel}) \, \delta [\omega(k_{\perp},k_{\parallel})-k_{\parallel}v_{\parallel}] \,\,\, ,
\end{equation}
where we have neglected the FLR effect discussed above.

For such a resonant acceleration process the force [Equation~(\ref{mir})] is the gradient of a potential.  Hence, by Equation~(\ref{tauaccreswave}), $\tau_{L}\propto 1/v^{3}$.  Since the turbulence is weak, the Lagrangian time scale should also be proportional to the Alfv\'en time $\tau_{A} = \lambda/V_A$, where $\lambda$ is the characteristic length scale, so that $\tau_{L}\sim (V_{A}/v)^{3} \, (\lambda/V_{A})$. Moreover, the magnitude of the force~(\ref{mir}) is $F\sim m_{e}v^{2} \, (1/\lambda) \, (B/B_{0})$. Hence we expect that

\begin{equation}\label{mirguess}
D(p)=\tau_{L} \, F^{2} \sim m_{e}V_{A}^{2} \, \left ( {1 \over \lambda} \right ) \, \left ( {B \over B_{0}} \right )^{2} \, p \,\,\, .
\end{equation}

The exact expression may be obtained as follows.  Using the expression~(\ref{mir}) for the magnetic mirror force and the dispersion relation $\omega=kV_{A}$ for fast-mode waves (where $V_A$ is the Alfv\'en speed), we obtain

\begin{equation}\label{mil}
D(p_{\parallel},p_{\perp})=\frac{\pi^{2} m_{e}^{2}}{B_{0}^{2}} \, v_{\perp}^{4}\int k_{\perp}^{2} \, dk_{\perp}\int dk_{\parallel} \, k_{\parallel}^{2} \, S_{B_{\parallel}}(k_{\perp},k_{\parallel}) \, \delta (kV_{A}-k_{\parallel}v_{\parallel}) \,\,\, ,
\end{equation}
with $S_{B_{\parallel}}(k_{\perp},k_{\parallel})$ the spectrum of the field strength fluctuations $B_{\parallel}$.  This diffusion coefficient may also, of course, be written in terms of the two independent variables $p$ and $\mu$. Assuming that efficient pitch-angle scattering maintains near-isotropy of the electron velocity distribution function, the momentum diffusion coefficient may then be averaged over $\mu$:

\begin{equation}
D(p)=\frac{1}{2}\int_{-1}^{1} d\mu \, D(\mu,p) \,\,\,  ,
\end{equation}
with the result \citep{mila,miller}

\begin{equation}\label{milert}
D(p)=\frac{1}{16} \, m_{e} \, V_{A}^{2} \, \frac{1}{\lambda} \, \left < \frac{B}{B_{0}} \right >^{2} \, p \,\,\, ,
\end{equation}
consistent with the estimate~(\ref{mirguess}).  Note that for this mechanism,

\begin{equation}\label{tauaccmirror}
D(p)\propto p ; \quad \tau_{\rm acc} \propto p \propto E^{1/2} \,\,\, .
\end{equation}

Transit-time acceleration by fast modes was also studied by \citet{si} who use the full expression for the diffusion tensor \citep{1992JPhG...18.1089J}. They show that for flat ($1<q\leq 2$) turbulence power spectra, the acceleration time scale increases as $\tau_{\rm acc}\propto E^{(3-q)/2}$, while for steep spectra ($2<q<6$) they recover $\tau_{\rm acc}\propto E^{1/2}$ at nonrelativistic energies, see Equations ($102$) and ($103$) in \citet{si}. \citet{yan} also discuss transit-time acceleration by fast modes in the low-$\beta$ plasma of the solar corona, they include a high-$k$ cutoff scale owing to collisionless damping of the fast waves resulting in $\tau_{\rm acc}\propto E^{1/2}$.

\subsection{The Parker equation -- adiabatic compression}\label{parker}

The interaction of particles with low-frequency compressible turbulence can also be modeled by the \citet{parker} equation, in which particles are accelerated by a compressive force. This force or the rate of particle momentum change follows an equation which is obtained from an adiabatic thermodynamic process

\begin{equation}\label{cp}
\frac{dp}{dt}=-\frac{1}{3} \, \nabla \cdot \mathbf{V} \, p \,\,\, .
\end{equation}
The evolution of the isotropic part of the distribution function is given by

\begin{equation}
\frac{\partial f}{\partial t}+\mathbf{V} \cdot \nabla f=\nabla \cdot \kappa \nabla f+\frac{1}{3} \, p \, \nabla \cdot \mathbf{V} \, \frac{\partial f}{\partial p} \,\,\, .
\end{equation}
Notice that spatial diffusion forms the basis of the Parker equation, such that in this framework stochastic acceleration is always non-resonant \citep{by,zh,j}. When the Peclet number is small $(VL/\kappa \ll 1)$ and when $V\ll \sqrt{\kappa/\tau}$, turbulent velocity fluctuations are unimportant for spatial transport, while still producing acceleration. However, in general the presence of turbulent velocity fluctuations $V$ enhances the spatial transport coefficient and, hence, the turbulent transport coefficient $\kappa_{t}$ is given by

\begin{equation}
\kappa_{t}=\kappa+\tau <V^{2}> \,.
\end{equation}
The momentum diffusion coefficient in the case of weak turbulence was written in, e.g., \citet{by} as

\begin{equation}
D(p) = {p^2 \over 9} \int d\mathbf{k} \ S(\mathbf{k}) \, \frac{(\kappa_{t} k^{2}+\tau_{c}^{-1})}{[\omega^{2}(\mathbf{k})+(\kappa_{t} k^{2}+\tau_{c}^{-1})^{2}]} \,\,\, ,
\end{equation}
with $\int d\mathbf{k} \, S(\mathbf{k}) = \, <(\nabla.\mathbf{V})^{2}> .$ [This result can be obtained by blending Equations~(\ref{trans_broad}) and~(\ref{turb_broad}).]

When the correlation time $\tau_{c}\rightarrow \infty$, the momentum diffusion coefficient becomes

\begin{equation}\label{nr}
D(p) = {p^2 \over 9} \int d\mathbf{k}  \, S(\mathbf{k}) \,\frac{ \kappa_{t} k^{2}}{[\omega^{2}(\mathbf{k})+(\kappa_{t}k^{2})^2]} \,\,\, .
\end{equation}
The spectral density $S(\mathbf{k})$ is in principle related to the spectrum of plasma density fluctuations, while the dispersion relation $\omega(\mathbf{k})$ may correspond to any low-frequency compressive modes, including an anisotropic spectrum of kinetic Alfv\'en waves \citep{2010PhPl...17f2308B}. Both fast and slow magnetosonic wave modes have been considered \citep{ptu,chan1,chan2,cho}.

\subsubsection{Example 1: Non-resonant acceleration by adiabatic compressions associated with fast-mode waves}\label{fast_waves}

In this case the dispersion relation $\omega(k)=kV_{A}$.  Assuming isotropic turbulence $S(\mathbf{k}) \, d\mathbf{k}=4\pi k^{2}S(k) \, dk$, it is found that in the strong diffusion limit $\kappa_t/\lambda \gg V_A $,

\begin{equation}
\frac{D(p)}{p^{2}} \rightarrow \left ( {\kappa_t \over V_A^2} \right ) \, <(\nabla \cdot \mathbf{V})^2> \,\,\, ,
\end{equation}
while in the small diffusion limit $\kappa_t/\lambda \ll V_A $,

\begin{equation}
\frac{D(p)}{p^{2}}  \rightarrow \left ( {\lambda^2 \over \kappa_t} \right )  \, <(\nabla \cdot \mathbf{V})^2> \,\,\,.
\end{equation}
Thus, for this mechanism, we obtain the relations

\begin{equation}\label{tauaccadiabatic}
D(p)\propto p^{2} ;  \quad \tau_{\rm acc} \propto {\rm constant} \,\,\, .
\end{equation}
Note that \citet{ptu} has discussed the case of a power-law spectrum, i.e., $S(k)\propto k^{q}$, and has concluded that that the scaling of $D$ with $p$ is insensitive to the value of the spectral index $q$.

\subsubsection{Example 2: Non-resonant acceleration by adiabatic compressions associated with strong turbulence}\label{strong_turbulence}

For the case of strong turbulence \citep{bf},

\begin{equation}
D(p) = {p^2 \over 9} \int d\omega \int d\mathbf{k} \, S(k,\omega) \,\frac{\kappa_{t} k^{2}}{[\omega^{2}+(\kappa_{t}k^{2})^{2}]} \,\,\, .
\end{equation}
In the weak diffusion limit ($\kappa_t \ll \lambda^2/\tau$)

\begin{equation}
\frac{D(p)}{p^{2}} \rightarrow \tau <(\nabla \cdot \mathbf{V})^2> \,\,\, ,
\end{equation}
while in the strong diffusion limit ($\kappa_t \gg \lambda^2/\tau$),

\begin{equation}
\frac{D(p)}{p^{2}}\rightarrow  \left ( {\lambda^2 \over \kappa_t} \right ) <(\nabla \cdot \mathbf{V})^2> \,\,\, .
\end{equation}
For both cases, therefore, we obtain the relationships

\begin{equation}\label{tauaccstrong}
D(p)\propto p^{2} ; \quad \tau_{\rm acc} \propto {\rm constant} \,\,\, .
\end{equation}
Most non-resonant acceleration models are based on the adiabatic compression relationship~(\ref{cp}). Note, however, that compressive waves can non-resonantly accelerate particles through both the parallel electric force and the magnetic mirror force.

\section{Energy Spectra}\label{energy_spectra}

In the examples discussed above, the momentum diffusion coefficient is a power law, i.e., $D(p)=D_{0} p^{\alpha}$. Therefore, in three dimensions, the distribution function $f(p,t)$ obeys the diffusion equation

\begin{equation}\label{eqdiffuse}
\frac{\partial f(p,t)}{\partial t}=\frac{1}{p^{2}} \, \frac{\partial }{\partial p} \left [ D_{0} \, p^{\alpha+2} \, \frac{\partial f(p,t)}{\partial p} \right ] \,\,\, ,
\end{equation}
with the normalization provided by $\int _{0}^{\infty} 4 \pi p^{2} \, f(p,t) \, dp=1$. Taking the initial condition as $f(p,0)=\delta(p)$, the general solution is the stretched exponential

\begin{equation}
f(p,t)=\frac{C}{4\pi} \, t^{3/(2-\alpha)} \, \exp \left [ \frac{-p^{2-\alpha}}{(2-\alpha)^{2}D_{0}t} \right ] \,\,\, ,
\end{equation}
with the constant $C=(2-\alpha)^{-(1+\alpha)/(2-\alpha)} \, D_{0}^{-3/(2-\alpha)} \, \Gamma[3/(2-\alpha)]$. This solution is invalid for $\alpha>2$ since $f$ becomes a monotonically increasing function of $p$. For arbitrary $\alpha$, a general solution of ($\ref{eqdiffuse}$) can be found with initial condition $f(p,0)=\delta(p-p_i)$, where $p_i$ is the initial momentum.  In units such that $p_i=1$, we obtain \citep{1986ApJ...308..929B,j}

\begin{equation}
1 \le \alpha < 2:
f(p,t)=\frac{p^{-(1+\alpha)/2}}{(2-\alpha) 4\pi D_{0}t} \, I_{l} \, \left [\frac{2p^{(2-\alpha)/2}}{(2-\alpha)^{2} \, D_{0}t} \right ] \, \exp \left [-\frac{1+p^{2-\alpha}}{(2-\alpha)^{2} \, D_{0}t} \right ] \,\,\, ,
\end{equation}
where $l=(1+\alpha)/(2-\alpha)$, and

\begin{equation}
\alpha > 2:
f(p,t)=\frac{p^{-(1+\alpha)/2}}{(\alpha-2) 4\pi D_{0}t} \, I_{l} \, \left [\frac{2p^{(2-\alpha)/2}}{(\alpha-2)^{2} \, D_{0}t} \right ] \, \exp \left [-\frac{1+p^{2-\alpha}}{(\alpha-2)^{2} \, D_{0}t} \right ] \,\,\, ,
\end{equation}
where $l=(\alpha+1)/(\alpha-2)$.  In the limit $t \rightarrow \infty$, the $\alpha > 2$ solution approaches the power-law form

\begin{equation}\label{large_time_limit}
\alpha > 2:
f(p,t \rightarrow \infty) \propto \left ( {1 \over D_{0}t} \right )^{2 \alpha -1 \over \alpha - 2} \, p^{-(1 + \alpha)} \,\,\, .
\end{equation}
The case $\alpha=2$ is special \citep{1962SvA.....6..317K}:

\begin{equation}
f(p,t)=\frac{1}{4\pi D_{0}t} \, p^{-3/2} \, \exp \left (-\frac{9}{4} \, D_{0}t \right ) \, \exp \left (-\frac{\ln^{2}p}{4D_{0}t} \right ) \,\,\, .
\end{equation}
These solutions generalize to the case where $D_{0}$ is a function of time by replacing the term $D_{0}t$ by $\int^{t} D_{0}(t') \, dt'$ in the previous expressions. Thus, power-law spectra can be obtained only for $\alpha >2$, in which case $f\sim p^{-(1+\alpha)}$; the case $\alpha=2$ leads to $f\sim p^{-3/2}$.

For continuous injection of particles in the acceleration region, i.e., when a source term $\delta(p-1)$ is added to the right side of the diffusion Equation~(\ref{eqdiffuse}), a {\it stationary} power-law solution may nevertheless result.  For such a stationary solution to exist, the momentum flux  $D_0 p^{\alpha+2} \, \partial f/\partial p$ must be constant, which requires that

\begin{equation}
f(p) \sim p^{-(1+\alpha)}; \quad \alpha > - 1 \,\,\, .
\end{equation}
Note that this has the same power-law spectral dependence on momentum as the (time-dependent) $\alpha > 2$ large-time limit solution~(\ref{large_time_limit}).

\section{Summary}\label{summary_sec}

To summarize, in the stochastic acceleration models discussed above the momentum diffusion coefficient $D(p)$ may be written in the equivalent forms

\begin{equation}\label{aa_summary}
D(p)=\int \int dt \, dx \, C(x,t) \, P(x,t) \, ; \qquad D(p) =  \int \int dk \, d\omega \, S(k,\omega) \, G(k,\omega) \,\,\, ,
\end{equation}
where $C(x,t)$ is the Eulerian correlation function of a force $F(x,t)$, $S(k,\omega)$ is the Fourier transform of $C(x,t)$, $P(x,t)$ is the solution of the spatial transport equation, and $G(k,\omega)$ is the Fourier transform of the associated Green's function. Given these general expressions, stochastic acceleration models are grouped as resonant, non-resonant or resonance broadened, depending on the form of $P(x,t)$ or, equivalently $G(k,\omega)$, which both describe the spatial transport of the particles in the acceleration region. For each case, we now summarize the pertinent equation for $P(x,t)$, its solution, and the resulting $G(k,\omega)$.

\begin{itemize}
\item {\it Resonant}

$$\frac{\partial P(x,t)}{\partial t}+v \, \frac{\partial P(x,t)}{\partial x}=0 \,\,\, ;$$
\begin{equation}
P(x,t)=\delta(x-vt) \,\,\, ; \quad D(p)=\pi \int \int dk \, d\omega \, S(k,\omega) \, \delta (\omega-kv) \,\,\, .
\end{equation}

\item
{\it Non-resonant}

$$\frac{\partial P(x,t)}{\partial t}=\kappa \, \frac{\partial^{2}P(x,t)}{\partial x^{2}} \,\,\, ;$$
\begin{equation}
P(x,t)=\frac{1}{\sqrt{4\pi \kappa t}} \,\, e^{-x^{2}/4\kappa t} \,\,\, ; \quad D(p)=\int \int dk \, d\omega \, S(k,\omega) \, \frac{\kappa k^{2}}{\omega^{2}+(\kappa k^{2})^{2}} \,\,\, .
\end{equation}

\item
{\it Resonant-broadened}

This combines elements of both the resonant and non-resonant cases:

$$\frac{\partial P(x,v,t)}{\partial t}+v \, \frac{\partial P(x,v,t)}{\partial x} = \kappa \, \frac{\partial^{2}P(x,t)}{\partial x^{2}} \,\,\, ;$$

\begin{equation}
P(x,t)=\frac{1}{\sqrt{4\pi \kappa t}}e^{-(x-vt)^{2}/4\kappa t} \,\,\, ; \quad D(p)=\int \int dk \, d\omega \, S(k,\omega) \, \frac{\kappa k^{2}}{(\omega-kv)^{2}+(\kappa k^{2})^2} \,\,\, .
\end{equation}

\end{itemize}

Finite Larmor-radius effects can be accounted for by writing the momentum diffusion coefficient as

\begin{equation}
D(p)=  \int \int d\mathbf{k} \, d\omega \,J^{2}_{0}(k_{\perp}\rho)
\, S(\mathbf{k},\omega) \, G(\mathbf{k},\omega) \,\,\, ; \qquad
G(\mathbf{k},\omega) =\int_0^\infty dt \,
<e^{i\mathbf{k}.\mathbf{X}(t)-i\omega t}> \,\,\, ,
\end{equation}
where $G(\mathbf{k},\omega)$ is now the propagator associated with the guiding-center $\mathbf{X}(t)$ of the particles.

The momentum dependence of $D(p)$ arises either because the accelerating force is itself momentum dependent ($S(k,\omega)$ depends on $p$) or because the spatial transport is momentum dependent ($G(k,\omega)$ depends on $p$). A typical dependence $D(p)$ is a power law

\begin{equation}
D(p)=D_{0} p^{\alpha} \,\,\, .
\end{equation}
Then, since $d<p^{2}>/dt=2D(p) = 2D_{0}p^{\alpha}$, it follows that

\begin{equation}
<p^{2}> \, \propto t^{2/(2-\alpha)} \,\,\, ,
\end{equation}
for $\alpha<2$.  For $\alpha > 2$, $<p^2>$ can attain infinite values within a finite period of time. Power-law distribution functions, solutions of the momentum diffusion equation are obtained only for $\alpha >2$, in which case $f\sim p^{-(1+\alpha)}$. When $\alpha<2$ the solutions are stretched exponentials. We also noticed that it may occur that the momentum diffusion coefficient is not existing because the integral involved in its expression is non-definite. We gave a specific example and dumb such acceleration mechanism fractional-order Fermi process.

All the above models (resonant/broadened/non-resonant) can apply equally well to describe stochastic acceleration in a force field which is \emph{fragmented}, or distributed in clumps, as is already known for the resonant interaction between particles and Langmuir waves \citep{1989SoPh..123..343M}. Noteworthy, in both resonant-broadened and non-resonant acceleration the spectrum can also be taken to be \emph{discrete} and even \emph{monochromatic}.

In general, the wavenumber $k$ and frequency $\omega$ are {\it a priori} independent quantities. However, in the case of acceleration by waves (e.g., MHD waves, plasma waves), these quantities are related by a dispersion relation $\omega = \omega(k)$. In such a case,

\begin{equation}
S(k,\omega) = S(k) \, \delta [\omega-\omega(k)] \,\,\, .
\end{equation}
For strong turbulence, the frequency spectrum of the force field is broadened around $\omega(k)$, so that

\begin{equation}
S(k,\omega) = \pi^{-1} S(k) \, \frac{\tau_{c}^{-1}}{[\omega-\omega(k)]^{2}+(\tau_{c}^{-1})^{2}} \,\,\, ,
\end{equation}
where $\tau_{c}$ is a turbulence correlation time scale. As remarked above, it is important to note that while broadening of the function $G(k,\omega)$ arises as a transport effect, broadening of the function $S(k,\omega)$ does not. However, the formal consequence is the same -- the introduction of an additional time scale in the stochastic acceleration model, either as a particle transport time scale $\nu_{d}^{-1} = (\kappa k^2)^{-1}$ or as a turbulence correlation time scale $\tau_{c}$.

As a rule of thumb, the momentum diffusion coefficient takes the form

\begin{equation}
D(p)=\tau_{L}<F^{2}> \,\,\, ,
\end{equation}
where the Lagrangian correlation time $\tau_{L}$ is a statistical property of the field in the frame comoving with the particles. In all cases, the acceleration efficiency, i.e., $<p^{2}>$ as a function of time, is estimated using quasilinear theory. Within quasilinear theory, the accelerating field acts as a perturbation on the particle motion which occurs at constant energy and is either integrable for resonant models or chaotic and diffusive for non-resonant models.

This general characterization of stochastic acceleration models allows us to use the dependence of the diffusion coefficient $D$ on the particle momentum $p$ to compare the acceleration times, acceleration rates, and energy spectra for a variety of acceleration models.  Specific expressions for the momentum dependence of the diffusion constant $D(p)$ and for the corresponding energy dependence of the acceleration time $\tau_{acc}(E)$ have been provided throughout the text.
%\begin{itemize}
%
%\item Equations~(\ref{tauaccres}) and~(\ref{tauaccreswave}) (for resonant acceleration -- Section~\ref{res_accn_models});
%\item Equation~(\ref{tauaccnonres}) (for non-resonant acceleration -- Section~\ref{non_res_accn_models});
%\item Equation~(\ref{tauaccfracfermi}) (for fractional-order Fermi acceleration -- Section~\ref{frac_order_fermi});
%\item Equation~(\ref{tauaccmirror}) (for acceleration in magnetic mirrors -- Section~\ref{accn_by_mirrors});
%\item Equation~(\ref{tauaccadiabatic}) (for acceleration by fast-wave adiabatic compressions -- Section~\ref{fast_waves}); and
%\item Equation~(\ref{tauaccstrong}) (for acceleration in strong turbulence -- Section~\ref{strong_turbulence}).
%\end{itemize}
These predicted scalings can be compared with quantities deduced from observation, and so to assess the viability of a given model. Applicable observations include the rise time of the radiation field (e.g., hard X-ray bremsstrahlung, gamma-rays) produced by the accelerated particles. Hard X-ray measurements also provide information on the {\it specific acceleration rate} $\nu_{specific}$ \citep[particles~s$^{-1}$ per particle;][]{2008AIPC.1039....3E}, a quantity which measures the fraction of the ambient particle population that suffers acceleration to energies equal to or greater than a prescribed energy. With a framework now in place to determine the diffusion coefficient for a general stochastic acceleration mechanism, the specific acceleration rate can be readily assessed and compared with values deduced from analysis of hard X-ray observations. This will be the subject of a future paper.

\appendix

\section{Generalization of Equations of Motion to Three Dimensions}

Applications of the results of Section~\ref{stoc_general} are straightforwardly generalizable to the following equations of motion in three dimensions:

\begin{equation}\label{s13d}
\dot{\mathbf{p}}=\mathbf{F}(\mathbf{x},t) \,\,\, , \qquad m\dot{\mathbf{x}}=\mathbf{p} \,\,\, .
\end{equation}
The evolution of the distribution function is given by a Fokker-Planck equation

\begin{equation}
\frac{\partial f(\mathbf{x},\mathbf{p},t)}{\partial t}+\frac{\mathbf {p}}{m} \cdot \nabla f(\mathbf{x},\mathbf{p},t)=\frac{\partial}{\partial p_{i}} \left [ D_{ij}(\mathbf{p}) \, \frac{\partial f(\mathbf{x},\mathbf{p},t)}{\partial {p_{j}}} \right ] \,\,\, .
\end{equation}
The expression for the momentum diffusion tensor is similar to Equation~(\ref{aa}) and becomes

\begin{equation}
D_{ij}(\mathbf{p})=\int_{0}^{\infty} dt\int d\mathbf{x} \, C_{ij}(\mathbf{x},t) \, P(\mathbf{x},t) \,\,\, ,
\end{equation}
where the Eulerian correlation tensor of the force field is defined by

\begin{equation}
C_{ij}(\mathbf{x},t) = \, <F_{i}(0,0) \, F_{j}(\mathbf{x},t)> \,\,\, ,
\end{equation}
while Equation~(\ref{bb}) becomes

\begin{equation}
D_{ij}(\mathbf{p}) =  \int \int d\mathbf{k} \, d\omega  \, S_{ij}(\mathbf{k},\omega) \, G(\mathbf{k},\omega) \,\,\, ,
\end{equation}
with definitions of $S_{ij}(\mathbf{k},\omega)$ and $G(\mathbf{k},\omega)$ similar to those in Equations~(\ref{sdefinition}) and~(\ref{prop_fourier}). A general decomposition of the Eulerian correlation tensor, assumed to be homogeneous and isotropic, is $C_{ij}(x,t)=A(x,t) \, \delta_{ij}+B(x,t) \, x_{i}x_{j}+G(x,t) \, \epsilon_{ijl} \, x_{l}$, where $\delta_{ij}$ is the Kronecker delta and $\epsilon_{ijl}$ is the permutation tensor. The symmetric part of this tensor is invariant relative to rotations and reflections. The antisymmetric part changes its sign under reflection and therefore is non-zero only for fields possessing helicity, i.e., when $\mathbf{F} \cdot \nabla \times \mathbf{F}\neq 0$. In a similar way, the general decomposition of the spectral tensor of the force field is $S_{ij}(k,t)=A(k,\omega) \, \delta_{ij}+B(k,\omega) \, k_{i}k_{j}/k^{2}+G(k,\omega) \, \epsilon_{ijl} \, k_{l}/k^{2}.$ The particular case of a potential field $\mathbf{F}=-\nabla \phi$, with $\mathbf{k}\times \mathbf{F}(\mathbf{k},\omega)=0$ (longitudinal
field), yields

\begin{equation}
S_{ij}(k,\omega)=S_{L}(k,\omega) \, \frac{k_{i}k_{j}}{k^{2}} \,\,\,
.
\end{equation}
The case of a solenoidal field $\nabla \cdot \mathbf{F}=0$, with $\mathbf{k} \cdot \mathbf{F}=0$ (transverse field), reads

\begin{equation}
S_{ij}(k,\omega)=S_{T}(k,\omega) \left ( \delta_{ij}-\frac{k_{i}k_{j}}{k^{2}} \right ) \,\,\, .
\end{equation}
Specific geometrical features of the field can be important in studying momentum and angular momentum injections into the particle motion \citep{1994PhyA..208..501C}. Here we only focus on the energy transfer from the field to the particles.

For example, these results can be used to calculate the momentum diffusion tensor resulting from resonant acceleration by a the force field which is the gradient of a potential $\mathbf{F}=-\nabla \phi$. In this case, $C_{ij}(x,t)=-\partial^{2}C_{\phi}(x,t)/{\partial x_{i}\partial x_{j}}$ and $S_{L}(k,\omega)=k^{2}S_{\phi}(k,\omega)$. We also assume that $\phi$ has a Gaussian correlation function. Then a calculation similar to that leading to Equation~(\ref{tauelltauresdim}) yields the following expression for the diffusion tensor

\begin{equation}
\frac{D_{ij}(\mathbf{p})}{D_{0}}=\frac{1}{[1+(p/p_{0})^{2}]^{3/2}} \, \frac{p_{i}p_{j}}{p^{2}}+\frac{1}{[1+(p/p_{0})^{2}]^{1/2}} \left (\delta_{ij}-\frac{p_{i}p_{j}}{p^{2}} \right ) \,\,\, ,
\end{equation}
with $D_{0}=\tau <\phi^{2}>/\lambda^{2}$. The  isotropic part of the distribution function obeys the diffusion equation (in $d$ dimensions):

\begin{equation}\label{diffueq}
\frac{\partial f(p,t)}{\partial t}=\frac{1}{p^{d-1}} \, \frac{\partial }{\partial p} \left [ {p^{d-1}} \, D(p) \, \frac{\partial f(p,t)}{\partial p} \right ] \,\,\, ,
\end{equation}
where

\begin{equation}\label{diffucoeff}
D(p) = \frac{D_0}{(1+p^{2}/p^{2}_{0})^{3/2}} \,\,\, .
\end{equation}
Notice that $D(p)\sim p^{-3}$ when $p\gg p_{0}$. This $D(p)\sim p^{-3}$ behavior, instead of $D(p)\sim p^{-1}$ [Equation~(\ref{tauaccres})], is a fundamental consequence of the force being the gradient of a potential.

\acknowledgments This work is partially supported by a STFC rolling grant. Financial support by the Leverhulme Trust (EPK) and by the European Commission through the "Radiosun" (PEOPLE-2011-IRSES-295272) and HESPE (FP7-SPACE-2010-263086) is gratefully acknowledged. AGE was supported by grant number NNX10AT78G from NASA's Heliospheric Physics Division.

\bibliographystyle{apj}
\bibliography{refacc}

\begin{thebibliography}{110}
\expandafter\ifx\csname natexlab\endcsname\relax\def\natexlab#1{#1}\fi

\bibitem[{{Achterberg}(1981)}]{acht}
{Achterberg}, A. 1981, \aap, 97, 259

\bibitem[{{Anastasiadis} {et~al.}(1997){Anastasiadis}, {Vlahos}, \&
  {Georgoulis}}]{an}
{Anastasiadis}, A., {Vlahos}, L., \& {Georgoulis}, M.~K. 1997, \apj, 489, 367

\bibitem[{{Antonucci} {et~al.}(1982){Antonucci}, {Gabriel}, {Acton},
  {Leibacher}, {Culhane}, {Rapley}, {Doyle}, {Machado}, \&
  {Orwig}}]{1982SoPh...78..107A}
{Antonucci}, E., {Gabriel}, A.~H., {Acton}, L.~W., {Leibacher}, J.~W.,
  {Culhane}, J.~L., {Rapley}, C.~G., {Doyle}, J.~G., {Machado}, M.~E., \&
  {Orwig}, L.~E. 1982, \solphys, 78, 107

\bibitem[{{Anzer}(1973)}]{1973SoPh...30..459A}
{Anzer}, U. 1973, \solphys, 30, 459

\bibitem[{{Arzner} \& {Vlahos}(2004)}]{arz}
{Arzner}, K., \& {Vlahos}, L. 2004, \apjl, 605, L69

\bibitem[{{Aschwanden}(2002)}]{as}
{Aschwanden}, M.~J. 2002, \ssr, 101, 1

\bibitem[{{Balescu}(2000)}]{2000EJPh...21..279B}
{Balescu}, R. 2000, European Journal of Physics, 21, 279

\bibitem[{{Barbosa}(1979)}]{1979ApJ...233..383B}
{Barbosa}, D.~D. 1979, \apj, 233, 383

\bibitem[{{Benka} \& {Holman}(1994)}]{ben}
{Benka}, S.~G., \& {Holman}, G.~D. 1994, \apj, 435, 469

\bibitem[{{Bian} \& {Browning}(2008)}]{bi}
{Bian}, N.~H., \& {Browning}, P.~K. 2008, \apjl, 687, L111

\bibitem[{{Bian} \& {Kontar}(2010)}]{2010PhPl...17f2308B}
{Bian}, N.~H., \& {Kontar}, E.~P. 2010, Physics of Plasmas, 17, 062308

\bibitem[{{Bian} {et~al.}(2010){Bian}, {Kontar}, \&
  {Brown}}]{2010A&A...519A.114B}
{Bian}, N.~H., {Kontar}, E.~P., \& {Brown}, J.~C. 2010, \aap, 519, A114

\bibitem[{{Bian} {et~al.}(2011){Bian}, {Kontar}, \&
  {MacKinnon}}]{2011A&A...535A..18B}
{Bian}, N.~H., {Kontar}, E.~P., \& {MacKinnon}, A.~L. 2011, \aap, 535, A18

\bibitem[{{Bieber} {et~al.}(1994){Bieber}, {Matthaeus}, {Smith}, {Wanner},
  {Kallenrode}, \& {Wibberenz}}]{1994ApJ...420..294B}
{Bieber}, J.~W., {Matthaeus}, W.~H., {Smith}, C.~W., {Wanner}, W.,
  {Kallenrode}, M.-B., \& {Wibberenz}, G. 1994, \apj, 420, 294

\bibitem[{{Borovsky} \& {Eilek}(1986)}]{1986ApJ...308..929B}
{Borovsky}, J.~E., \& {Eilek}, J.~A. 1986, \apj, 308, 929

\bibitem[{{Brown} {et~al.}(2009){Brown}, {Turkmani}, {Kontar}, {MacKinnon}, \&
  {Vlahos}}]{2009A&A...508..993B}
{Brown}, J.~C., {Turkmani}, R., {Kontar}, E.~P., {MacKinnon}, A.~L., \&
  {Vlahos}, L. 2009, \aap, 508, 993

\bibitem[{{Browning} {et~al.}(2010){Browning}, {Dalla}, {Peters}, \&
  {Smith}}]{2010A&A...520A.105B}
{Browning}, P.~K., {Dalla}, S., {Peters}, D., \& {Smith}, J. 2010, \aap, 520,
  A105

\bibitem[{{Bykov} \& {Fleishman}(2009{\natexlab{a}})}]{2009ApJ...692L..45B}
{Bykov}, A.~M., \& {Fleishman}, G.~D. 2009{\natexlab{a}}, \apjl, 692, L45

\bibitem[{{Bykov} \& {Fleishman}(2009{\natexlab{b}})}]{bf}
---. 2009{\natexlab{b}}, \apjl, 692, L45

\bibitem[{{Bykov} \& {Toptygin}(1993{\natexlab{a}})}]{1993PhyU...36.1020B}
{Bykov}, A.~M., \& {Toptygin}, I. 1993{\natexlab{a}}, Physics Uspekhi, 36, 1020

\bibitem[{{Bykov} \& {Toptygin}(1993{\natexlab{b}})}]{by}
---. 1993{\natexlab{b}}, Physics Uspekhi, 36, 1020

\bibitem[{{Cairns} \& {McMillan}(2005)}]{2005PhPl...12j2110C}
{Cairns}, I.~H., \& {McMillan}, B.~F. 2005, Physics of Plasmas, 12, 102110

\bibitem[{{Cargill} {et~al.}(2006){Cargill}, {Vlahos}, {Turkmani}, {Galsgaard},
  \& {Isliker}}]{car}
{Cargill}, P.~J., {Vlahos}, L., {Turkmani}, R., {Galsgaard}, K., \& {Isliker},
  H. 2006, \ssr, 124, 249

\bibitem[{{Casanova} \& {Schlickeiser}(2012)}]{2012ApJ...745..153C}
{Casanova}, S., \& {Schlickeiser}, R. 2012, \apj, 745, 153

\bibitem[{{Chandran}(2003)}]{chan1}
{Chandran}, B.~D.~G. 2003, \apj, 599, 1426

\bibitem[{{Chandran} \& {Maron}(2004)}]{chan2}
{Chandran}, B.~D.~G., \& {Maron}, J.~L. 2004, \apj, 603, 23

\bibitem[{{Chechkin} {et~al.}(1994){Chechkin}, {Tur}, \&
  {Yanovsky}}]{1994PhyA..208..501C}
{Chechkin}, A.~V., {Tur}, A.~V., \& {Yanovsky}, V.~V. 1994, Physica A
  Statistical Mechanics and its Applications, 208, 501

\bibitem[{{Chen} \& {Petrosian}(2012)}]{2012ApJ...748...33C}
{Chen}, Q., \& {Petrosian}, V. 2012, \apj, 748, 33

\bibitem[{{Cho} \& {Lazarian}(2006)}]{cho}
{Cho}, J., \& {Lazarian}, A. 2006, \apj, 638, 811

\bibitem[{{Corrsin}(1959)}]{cor}
{Corrsin}, S. 1959, Advances in Geophysics, 6, 441

\bibitem[{{Dreicer}(1959)}]{1959PhRv..115..238D}
{Dreicer}, H. 1959, Physical Review, 115, 238

\bibitem[{{Dupree}(1966)}]{dup}
{Dupree}, T.~H. 1966, Physics of Fluids, 9, 1773

\bibitem[{{Emslie} \& {H\'enoux}(1995)}]{1995ApJ...446..371E}
{Emslie}, A.~G., \& {H\'enoux}, J.-C. 1995, \apj, 446, 371

\bibitem[{{Emslie} {et~al.}(2008){Emslie}, {Hurford}, {Kontar}, {Massone},
  {Piana}, {Prato}, \& {Xu}}]{2008AIPC.1039....3E}
{Emslie}, A.~G., {Hurford}, G.~J., {Kontar}, E.~P., {Massone}, A.~M., {Piana},
  M., {Prato}, M., \& {Xu}, Y. 2008, in American Institute of Physics
  Conference Series, Vol. 1039, American Institute of Physics Conference
  Series, ed. {G.~Li, Q.~Hu, O.~Verkhoglyadova, G.~P.~Zank, R.~P.~Lin, \&
  J.~Luhmann }, 3--10

\bibitem[{{Escande} \& {Elskens}(2003)}]{2003PPCF...45A.115E}
{Escande}, D.~F., \& {Elskens}, Y. 2003, Plasma Physics and Controlled Fusion,
  45, A260000

\bibitem[{{Fermi}(1949)}]{1949PhRv...75.1169F}
{Fermi}, E. 1949, Physical Review, 75, 1169

\bibitem[{{Fludra} {et~al.}(1989){Fludra}, {Bentley}, {Lemen}, {Jakimiec}, \&
  {Sylwester}}]{1989ApJ...344..991F}
{Fludra}, A., {Bentley}, R.~D., {Lemen}, J.~R., {Jakimiec}, J., \& {Sylwester},
  J. 1989, \apj, 344, 991

\bibitem[{{Frost} \& {Dennis}(1971)}]{1971ApJ...165..655F}
{Frost}, K.~J., \& {Dennis}, B.~R. 1971, \apj, 165, 655

\bibitem[{{Giacalone} \& {K{\'o}ta}(2006)}]{2006SSRv..124..277G}
{Giacalone}, J., \& {K{\'o}ta}, J. 2006, \ssr, 124, 277

\bibitem[{{Gisler} \& {Lemons}(1990)}]{1990JGR....9514925G}
{Gisler}, G., \& {Lemons}, D. 1990, \jgr, 95, 14925

\bibitem[{{Goldreich} \& {Sridhar}(1995)}]{1995ApJ...438..763G}
{Goldreich}, P., \& {Sridhar}, S. 1995, \apj, 438, 763

\bibitem[{{Gordovskyy} \& {Browning}(2011)}]{2011ApJ...729..101G}
{Gordovskyy}, M., \& {Browning}, P.~K. 2011, \apj, 729, 101

\bibitem[{{Grady} \& {Neukirch}(2009)}]{2009A&A...508.1461G}
{Grady}, K.~J., \& {Neukirch}, T. 2009, \aap, 508, 1461

\bibitem[{{Hall} \& {Sturrock}(1967)}]{1967PhFl...10.2620H}
{Hall}, D.~E., \& {Sturrock}, P.~A. 1967, Physics of Fluids, 10, 2620

\bibitem[{{Holman} {et~al.}(2011){Holman}, {Aschwanden}, {Aurass}, {Battaglia},
  {Grigis}, {Kontar}, {Liu}, {Saint-Hilaire}, \&
  {Zharkova}}]{2011SSRv..159..107H}
{Holman}, G.~D., {Aschwanden}, M.~J., {Aurass}, H., {Battaglia}, M., {Grigis},
  P.~C., {Kontar}, E.~P., {Liu}, W., {Saint-Hilaire}, P., \& {Zharkova}, V.~V.
  2011, \ssr, 159, 107

\bibitem[{{Howes} {et~al.}(2006){Howes}, {Cowley}, {Dorland}, {Hammett},
  {Quataert}, \& {Schekochihin}}]{2006ApJ...651..590H}
{Howes}, G.~G., {Cowley}, S.~C., {Dorland}, W., {Hammett}, G.~W., {Quataert},
  E., \& {Schekochihin}, A.~A. 2006, \apj, 651, 590

\bibitem[{{Hoyng}(1977)}]{1977A&A....55...23H}
{Hoyng}, P. 1977, \aap, 55, 23

\bibitem[{{Jaekel} \& {Schlickeiser}(1992)}]{1992JPhG...18.1089J}
{Jaekel}, U., \& {Schlickeiser}, R. 1992, Journal of Physics G Nuclear Physics,
  18, 1089

\bibitem[{{Jokipii} {et~al.}(2003){Jokipii}, {Giacalone}, \&
  {Kota}}]{2003ICRC....6.3685J}
{Jokipii}, J.~R., {Giacalone}, J., \& {Kota}, J. 2003, in International Cosmic
  Ray Conference, Vol.~6, International Cosmic Ray Conference, 3685

\bibitem[{{Jokipii} \& {Lee}(2010)}]{j}
{Jokipii}, J.~R., \& {Lee}, M.~A. 2010, \apj, 713, 475

\bibitem[{{Kardashev}(1962)}]{1962SvA.....6..317K}
{Kardashev}, N.~S. 1962, \sovast, 6, 317

\bibitem[{{Karlick{\'y}} \& {Kosugi}(2004)}]{karl}
{Karlick{\'y}}, M., \& {Kosugi}, T. 2004, \aap, 419, 1159

\bibitem[{{Kaufman}(1972)}]{1972JPlPh...8....1K}
{Kaufman}, A.~N. 1972, Journal of Plasma Physics, 8, 1

\bibitem[{{Kennel} \& {Engelmann}(1966)}]{1966PhFl....9.2377K}
{Kennel}, C.~F., \& {Engelmann}, F. 1966, Physics of Fluids, 9, 2377

\bibitem[{{Kontar} {et~al.}(2011{\natexlab{a}}){Kontar}, {Brown}, {Emslie},
  {Hajdas}, {Holman}, {Hurford}, {Ka{\v s}parov{\'a}}, {Mallik}, {Massone},
  {McConnell}, {Piana}, {Prato}, {Schmahl}, \&
  {Suarez-Garcia}}]{2011SSRv..159..301K}
{Kontar}, E.~P., {Brown}, J.~C., {Emslie}, A.~G., {Hajdas}, W., {Holman},
  G.~D., {Hurford}, G.~J., {Ka{\v s}parov{\'a}}, J., {Mallik}, P.~C.~V.,
  {Massone}, A.~M., {McConnell}, M.~L., {Piana}, M., {Prato}, M., {Schmahl},
  E.~J., \& {Suarez-Garcia}, E. 2011{\natexlab{a}}, \ssr, 159, 301

\bibitem[{{Kontar} {et~al.}(2011{\natexlab{b}}){Kontar}, {Hannah}, \&
  {Bian}}]{2011ApJ...730L..22K}
{Kontar}, E.~P., {Hannah}, I.~G., \& {Bian}, N.~H. 2011{\natexlab{b}}, \apjl,
  730, L22

\bibitem[{{Kontar} \& {Mel'Nik}(2003)}]{2003PhPl...10.2732K}
{Kontar}, E.~P., \& {Mel'Nik}, V.~N. 2003, Physics of Plasmas, 10, 2732

\bibitem[{{Kontar} {et~al.}(2012){Kontar}, {Ratcliffe}, \&
  {Bian}}]{2012A&A...539A..43K}
{Kontar}, E.~P., {Ratcliffe}, H., \& {Bian}, N.~H. 2012, \aap, 539, A43

\bibitem[{{Krommes}(2002)}]{2002PhR...360....1K}
{Krommes}, J.~A. 2002, \physrep, 360, 1

\bibitem[{{Kubo}(1963)}]{1963JMP.....4..174K}
{Kubo}, R. 1963, Journal of Mathematical Physics, 4, 174

\bibitem[{{Kuijpers} {et~al.}(1981){Kuijpers}, {van der Post}, \&
  {Slottje}}]{1981A&A...103..331K}
{Kuijpers}, J., {van der Post}, P., \& {Slottje}, C. 1981, \aap, 103, 331

\bibitem[{{Kulsrud} \& {Ferrari}(1971)}]{kul}
{Kulsrud}, R.~M., \& {Ferrari}, A. 1971, \apss, 12, 302

\bibitem[{{Litvinenko}(1996)}]{li1}
{Litvinenko}, Y.~E. 1996, \apj, 462, 997

\bibitem[{{Litvinenko}(2000)}]{li2}
---. 2000, \solphys, 194, 327

\bibitem[{{Melrose}(1969)}]{1969Ap&SS...4..143M}
{Melrose}, D.~B. 1969, \apss, 4, 143

\bibitem[{{Melrose}(1994)}]{mel}
---. 1994, \apjs, 90, 623

\bibitem[{{Melrose}(2012)}]{2012ApJ...749...58M}
---. 2012, \apj, 749, 58

\bibitem[{{Melrose} \& {Cramer}(1989)}]{1989SoPh..123..343M}
{Melrose}, D.~B., \& {Cramer}, N.~F. 1989, \solphys, 123, 343

\bibitem[{{Miller} {et~al.}(1997){Miller}, {Cargill}, {Emslie}, {Holman},
  {Dennis}, {LaRosa}, {Winglee}, {Benka}, \& {Tsuneta}}]{miller}
{Miller}, J.~A., {Cargill}, P.~J., {Emslie}, A.~G., {Holman}, G.~D., {Dennis},
  B.~R., {LaRosa}, T.~N., {Winglee}, R.~M., {Benka}, S.~G., \& {Tsuneta}, S.
  1997, \jgr, 102, 14631

\bibitem[{{Miller} {et~al.}(1996){Miller}, {Larosa}, \& {Moore}}]{mila}
{Miller}, J.~A., {Larosa}, T.~N., \& {Moore}, R.~L. 1996, \apj, 461, 445

\bibitem[{{Neuer} \& {Spatschek}(2006)}]{2006PhRvE..73b6404N}
{Neuer}, M., \& {Spatschek}, K.~H. 2006, \pre, 73, 026404

\bibitem[{{Obukhov}(1959)}]{ob}
{Obukhov}, A.~M. 1959, Advances in Geophysics, 6, 113

\bibitem[{{Parker}(1965)}]{parker}
{Parker}, E.~N. 1965, \planss, 13, 9

\bibitem[{{Petkaki} \& {MacKinnon}(2007)}]{2007A&A...472..623P}
{Petkaki}, P., \& {MacKinnon}, A.~L. 2007, \aap, 472, 623

\bibitem[{{Petrosian} \& {Bykov}(2008)}]{2008SSRv..134..207P}
{Petrosian}, V., \& {Bykov}, A.~M. 2008, \ssr, 134, 207

\bibitem[{{Petrosian} {et~al.}(1994){Petrosian}, {McTiernan}, \&
  {Marschhauser}}]{1994ApJ...434..747P}
{Petrosian}, V., {McTiernan}, J.~M., \& {Marschhauser}, H. 1994, \apj, 434, 747

\bibitem[{{Press} {et~al.}(1992){Press}, {Teukolsky}, {Vetterling}, \&
  {Flannery}}]{1992nrca.book.....P}
{Press}, W.~H., {Teukolsky}, S.~A., {Vetterling}, W.~T., \& {Flannery}, B.~P.
  1992, {Numerical recipes in C. The art of scientific computing}

\bibitem[{{Ptuskin}(1988)}]{ptu}
{Ptuskin}, V.~S. 1988, Soviet Astronomy Letters, 14, 255

\bibitem[{{Pustilnik}(1975)}]{1975AZh....52..316P}
{Pustilnik}, L.~A. 1975, \azh, 52, 316

\bibitem[{{Ragot} \& {Schlickeiser}(1998)}]{1998A&A...331.1066R}
{Ragot}, B.~R., \& {Schlickeiser}, R. 1998, \aap, 331, 1066

\bibitem[{{Rudakov} \& {Tsytovich}(1971)}]{rud}
{Rudakov}, L.~I., \& {Tsytovich}, V.~N. 1971, Plasma Physics, 13, 213

\bibitem[{{Sagdeev} \& {Galeev}(1969)}]{sag}
{Sagdeev}, R.~Z., \& {Galeev}, A.~A. 1969, {Nonlinear Plasma Theory}, ed.
  {Sagdeev, R.~Z.~\& Galeev, A.~A.}

\bibitem[{{Schekochihin} {et~al.}(2009){Schekochihin}, {Cowley}, {Dorland},
  {Hammett}, {Howes}, {Quataert}, \& {Tatsuno}}]{2009ApJS..182..310S}
{Schekochihin}, A.~A., {Cowley}, S.~C., {Dorland}, W., {Hammett}, G.~W.,
  {Howes}, G.~G., {Quataert}, E., \& {Tatsuno}, T. 2009, \apjs, 182, 310

\bibitem[{{Schlickeiser}(2011)}]{2011ApJ...732...96S}
{Schlickeiser}, R. 2011, \apj, 732, 96

\bibitem[{{Schlickeiser} \& {Achatz}(1993)}]{1993JPlPh..49...63S}
{Schlickeiser}, R., \& {Achatz}, U. 1993, Journal of Plasma Physics, 49, 63

\bibitem[{{Schlickeiser} \& {Miller}(1998)}]{si}
{Schlickeiser}, R., \& {Miller}, J.~A. 1998, \apj, 492, 352

\bibitem[{{Skilling}(1975)}]{1975MNRAS.172..557S}
{Skilling}, J. 1975, \mnras, 172, 557

\bibitem[{{Somov}(1986)}]{1986A&A...163..210S}
{Somov}, B.~V. 1986, \aap, 163, 210

\bibitem[{{Somov} \& {Bogachev}(2003)}]{som2}
{Somov}, B.~V., \& {Bogachev}, S.~A. 2003, Astronomy Letters, 29, 621

\bibitem[{{Somov} \& {Kosugi}(1997)}]{som1}
{Somov}, B.~V., \& {Kosugi}, T. 1997, \apj, 485, 859

\bibitem[{{Sturrock}(1966)}]{stu}
{Sturrock}, P.~A. 1966, Physical Review, 141, 186

\bibitem[{{Sweet}(1969)}]{1969ARA&A...7..149S}
{Sweet}, P.~A. 1969, \araa, 7, 149

\bibitem[{{Syrovatskii}(1966)}]{1966AZh....43..340S}
{Syrovatskii}, S.~I. 1966, \azh, 43, 340

\bibitem[{{Takakura}(1988)}]{1988SoPh..115..149T}
{Takakura}, T. 1988, \solphys, 115, 149

\bibitem[{{Tautz} \& {Shalchi}(2010)}]{tau}
{Tautz}, R.~C., \& {Shalchi}, A. 2010, Physics of Plasmas, 17, 122313

\bibitem[{{Taylor}(1922)}]{taylor}
{Taylor}, G.~I. 1922, Proceedings of the London Mathematical Society, 20, 196

\bibitem[{Thomson \& Benford(1972)}]{benf}
Thomson, J.~J., \& Benford, G. 1972, Phys. Rev. Lett., 28, 590

\bibitem[{{Tsytovich}(1966)}]{1966SvPhU...9..370T}
{Tsytovich}, V.~N. 1966, Soviet Physics Uspekhi, 9, 370

\bibitem[{{Tsytovich}(1995)}]{1995lnlp.book.....T}
---. 1995, {Lectures on Non-linear Plasma Kinetics}, ed. {Tsytovich, V.~N.~\&
  ter Haar, D.}

\bibitem[{{Turkmani} {et~al.}(2005){Turkmani}, {Vlahos}, {Galsgaard},
  {Cargill}, \& {Isliker}}]{turkmanietal05}
{Turkmani}, R., {Vlahos}, L., {Galsgaard}, K., {Cargill}, P.~J., \& {Isliker},
  H. 2005, \apjl, 620, L59

\bibitem[{{Vlad} {et~al.}(2004){Vlad}, {Spineanu}, {Misguich}, {Reuss},
  {Balescu}, {Itoh}, \& {Itoh}}]{2004PPCF...46.1051V}
{Vlad}, M., {Spineanu}, F., {Misguich}, J.~H., {Reuss}, J.-D., {Balescu}, R.,
  {Itoh}, K., \& {Itoh}, S.-I. 2004, Plasma Physics and Controlled Fusion, 46,
  1051

\bibitem[{{Wang} {et~al.}(1995){Wang}, {Vlad}, {vanden Eijnden}, {Spineanu},
  {Misguich}, \& {Balescu}}]{1995PhRvE..51.4844W}
{Wang}, H.-D., {Vlad}, M., {vanden Eijnden}, E., {Spineanu}, F., {Misguich},
  J.~H., \& {Balescu}, R. 1995, \pre, 51, 4844

\bibitem[{{Webb} {et~al.}(2006){Webb}, {Zank}, {Kaghashvili}, \& {le
  Roux}}]{2006ApJ...651..211W}
{Webb}, G.~M., {Zank}, G.~P., {Kaghashvili}, E.~K., \& {le Roux}, J.~A. 2006,
  \apj, 651, 211

\bibitem[{{Weinstock}(1968)}]{wein}
{Weinstock}, J. 1968, Physics of Fluids, 11, 1977

\bibitem[{{Wentzel}(1974)}]{1974ARA&A..12...71W}
{Wentzel}, D.~G. 1974, \araa, 12, 71

\bibitem[{{Yan} {et~al.}(2008){Yan}, {Lazarian}, \& {Petrosian}}]{yan}
{Yan}, H., {Lazarian}, A., \& {Petrosian}, V. 2008, \apj, 684, 1461

\bibitem[{{Zhang}(2010)}]{2010JGRA..11512102Z}
{Zhang}, M. 2010, Journal of Geophysical Research (Space Physics), 115, 12102

\bibitem[{{Zhang} \& {Lee}(2011)}]{zh}
{Zhang}, M., \& {Lee}, M.~A. 2011, \ssr, 25

\bibitem[{{Zharkova} {et~al.}(2011){Zharkova}, {Arzner}, {Benz}, {Browning},
  {Dauphin}, {Emslie}, {Fletcher}, {Kontar}, {Mann}, {Onofri}, {Petrosian},
  {Turkmani}, {Vilmer}, \& {Vlahos}}]{2011SSRv..159..357Z}
{Zharkova}, V.~V., {Arzner}, K., {Benz}, A.~O., {Browning}, P., {Dauphin}, C.,
  {Emslie}, A.~G., {Fletcher}, L., {Kontar}, E.~P., {Mann}, G., {Onofri}, M.,
  {Petrosian}, V., {Turkmani}, R., {Vilmer}, N., \& {Vlahos}, L. 2011, \ssr,
  159, 357

\bibitem[{{Zharkova} \& {Gordovskyy}(2005)}]{2005MNRAS.356.1107Z}
{Zharkova}, V.~V., \& {Gordovskyy}, M. 2005, \mnras, 356, 1107

\end{thebibliography}

\end{document}